\documentclass[aps,pra,twocolumn,showpacs,superscriptaddress]{revtex4-2} 

\usepackage{graphicx}
\usepackage{xr-hyper}  
\usepackage[colorlinks,citecolor=blue,urlcolor=blue,linkcolor=blue]{hyperref}
\usepackage{siunitx}	
\usepackage{amsmath}
\usepackage{amsthm}
\usepackage{amssymb}
\usepackage{amsthm}
\usepackage{braket}
\usepackage{booktabs}
\usepackage{footmisc}
\usepackage{mathrsfs}
\usepackage{cleveref}
\usepackage{dsfont}
\usepackage{subfigure}
\usepackage{multirow}
\usepackage{enumerate}
\usepackage{lineno}

\newcommand{\bracket}[2]{\ensuremath{\langle#1 \vphantom{#2}| #2\vphantom{#1}\rangle}}

\newcommand{\ketbra}[2]{\ensuremath{|#1 \vphantom{#2}\rangle \langle #2\vphantom{#1}|}}

\newcommand{\unit}{\mathds{1}}

\newcommand{\sgn}{\mathrm{sgn}}

\newcommand{\tr}[1]{\mathrm{Tr}\left(#1\right)}

\newcommand{\abs}[1]{\left| #1 \right|} 

\newcommand{\im}{\mathrm{i}}

\begin{document}
\title{First-detection-time statistics in many-body quantum transport}

\author{Christoph Dittel}
\email{christoph.dittel@physik.uni-freiburg.de}
\affiliation{Physikalisches Institut, Albert-Ludwigs-Universit{\"a}t Freiburg, Hermann-Herder-Straße 3, 79104 Freiburg, Germany}
\affiliation{EUCOR Centre for Quantum Science and Quantum Computing, Albert-Ludwigs-Universität Freiburg, Hermann-Herder-Straße 3, 79104 Freiburg, Germany}
\affiliation{Freiburg Institute for Advanced Studies, Albert-Ludwigs-Universität Freiburg, Albertstraße 19, 79104 Freiburg, Germany}

\author{Niklas Neubrand}
\affiliation{Physikalisches Institut, Albert-Ludwigs-Universit{\"a}t Freiburg, Hermann-Herder-Straße 3, 79104 Freiburg, Germany}

\author{Felix Thiel}
\email{thiel@posteo.de}
\affiliation{Physikalisches Institut, Albert-Ludwigs-Universit{\"a}t Freiburg, Hermann-Herder-Straße 3, 79104 Freiburg, Germany}

\author{Andreas Buchleitner}
\affiliation{Physikalisches Institut, Albert-Ludwigs-Universit{\"a}t Freiburg, Hermann-Herder-Straße 3, 79104 Freiburg, Germany}
\affiliation{EUCOR Centre for Quantum Science and Quantum Computing, Albert-Ludwigs-Universität Freiburg, Hermann-Herder-Straße 3, 79104 Freiburg, Germany}

\date{\today}

\begin{abstract}
We study the transport of many partially distinguishable and possibly interacting particles under the action of repeated projective measurements on a target space and investigate how the particles' interference affects the mean first detection time. We contrast the detection of exactly $n$ versus at least $n$ particles, explain divergences in the mean first detection time through spectral properties of the generating evolution operator, and illustrate our findings by an example. 
\end{abstract}

\maketitle

The traveling time of an object through a potential landscape not only is an important quantity in classical systems, it likewise plays a central role in quantum transport problems, e.g., as a quantifier of the transport's efficiency \cite{Farhi-QC-1998,Ambainis-OD-2001}. However, while the detection time of a continuously moving object at a target space is well defined in classical mechanics, in standard quantum mechanics it is not, since time is no observable and thus cannot be measured directly. An increasingly popular approach to circumvent this problem is to perform repeated projective measurements on the target space \cite{Varbanov-HT-2008,Gruenbaum-RD-2013} such that the first detection time can be well defined as the mean time until the object is detected for the first time.  Intuitively, this detection scheme strongly depends on the time intervals between consecutive measurements and results in the quantum Zeno effect \cite{Degasperis-DL-1974} in the limit of small time steps compared to typical evolution times.

A natural choice is to consider equidistant time steps (called sampling times $\tau$) as in investigations of the transport of pure single-particle states along diverse tight-binding lattices \cite{Dhar-QT-2015,Dhar-DQ-2015,Friedman-QR-2016,Friedman-QW-2017,Thiel-FD-2018,Yin-LF-2019,Liu-QW-2020,Thiel-US-2020,Thiel-DS-2020}. \textit{Inter alia}, it was observed that the mean first-detection time diverges for particular, so-called resonant sampling times \cite{Friedman-QR-2016,Friedman-QW-2017,Yin-LF-2019,Liu-QW-2020,Thiel-DS-2020}, which disappear if the sampling times are chosen randomly \cite{Kessler-FD-2021}. While the appearance of these resonances was related to a classical electrostatic problem \cite{Gruenbaum-RD-2013,Liu-QW-2020}, a full spectral understanding remains desirable.

On the other hand, for the transport of many particles, promising applications, such as universal quantum computation \cite{OBrien-OQ-2007,Childs-UC-2013}, spurred much interest in exploiting particle indistinguishability \cite{Tichy-ZT-2010,Tichy-SP-2015,Shchesnovich-PI-2015,Dittel-AI-2019,Dittel-WP-2021,Minke-CF-2021}, in both interacting \cite{Lahini-QW-2012,Ahlbrecht-MB-2012,Preiss-SC-2015,Cai-MQ-2021} and noninteracting \cite{Poulios-QW-2014,Crespi-PS-2015,Ehrhardt-EC-2021} systems. The question therefore arises how stroboscopic measurements affect the transport properties in these many-body systems. This entails a series of new problems, e.g., how partial particle distinguishability or stroboscopic measurements with particle number resolution affect the mean first-detection time.

In this work we address these questions. By expanding the formalism so far available, we lift the concept of stroboscopic measurements to the realm of many partially distinguishable particles. To this end, we provide a general description of the first-detection-time statistics in the density operator formalism, valid independently of the exact physical scenario. This allows for the consideration of mixed states, an essential ingredient to treat partial particle distinguishability \cite{Dittel-AI-2019,Dittel-WP-2021,Minke-CF-2021,Brunner-MB-2022}. In this general formalism, we additionally address the identification of resonant sampling times through a spectral approach. In particular, we relate divergences of the mean first-detection time to the spectral properties of the so-called survival operators \cite{Dhar-QT-2015,Dhar-DQ-2015} whose expectation values account for the full first-detection-time statistics. 

We then apply our formalism to the transport of $N$ partially distinguishable, interfering particles, on a finite-dimensional Hilbert space. For a general number of $N$ particles, with transport properties generated by an arbitrary many-body unitary evolution operator (describing any conceivable hopping dynamics and possibly interacting particles), we discuss how distinguishability affects the first-detection-time statistics and compare the stroboscopic measurements corresponding to the detection of exactly $n\leq N$ and at least $n$ particles on a single target site. In particular, for the detection of all $N$ particles on a single target site, we find that the first-detection probabilities are proportional to the particle exchange symmetry of the initial state and that the first-detection time is independent of the particles' distinguishability. As an example, underpinning our analytical results, we finally provide numerical calculations for the mean first-detection time in the transport of two noninteracting, partially distinguishable particles on one-dimensional linear lattices of variable lengths. 

The paper is structured as follows. In Sec.~\ref{sec:StroMeas} we provide the density operator formalism for the stroboscopic measurement and in Sec.~\ref{sec:SurvOp} we further show that the first-detection-time statistics are fully determined by the expectation values of the survival operators. Section~\ref{sec:Trap} introduces the so-called trapped subspace with the help of which we determine the divergences of the mean first-detection time in Sec.~\ref{sec:divergence}. These resonances are shown to be anchored to the so-called degenerate subspace in Sec.~\ref{sec:degenerate}. In Sec.~\ref{sec:MPGenForm} we then apply our formalism to the transport of many partially distinguishable particles and discuss different types of stroboscopic measurements. We present a numerical illustration for the transport of two partially distinguishable particles on linear lattices in Sec.~\ref{sec:numerics}. We summarize in Sec.~\ref{sec:Conclusion}. For the sake of readability, all detailed proofs are deferred to the Appendixes.

\section{First detection time statistics}
\subsection{Stroboscopic measurement}\label{sec:StroMeas}
Let us start with providing a general description of stroboscopic measurements \cite{Dhar-QT-2015,Dhar-DQ-2015,Friedman-QR-2016,Friedman-QW-2017,Thiel-FD-2018,Yin-LF-2019,Liu-QW-2020,Thiel-US-2020,Thiel-DS-2020,Kessler-FD-2021} in the density operator formalism. To this end, consider a general quantum state (irrespective of the exact physical scenario) described by the density operator $\rho$ and suppose that it is initially prepared in the subspace $\mathscr{H}_\parallel$ of the total Hilbert space $\mathscr{H}=\mathscr{H}_\parallel \oplus \mathscr{H}_\perp$. We consider its coherent evolution generated by the Hamiltonian $\mathcal{H}$ and ask for the time delay to successfully detect the state in the \textit{detection subspace} $\mathscr{H}_\perp$ for the first time. Our strategy is to perform repetitive measurements after equidistant \textit{sampling times} $\tau$, according to the binary projective-valued measurement $\{\mathcal{P}_\perp,\mathcal{P}_\parallel\}$, where $\mathcal{P}_\perp$ and $\mathcal{P}_\parallel=\unit - \mathcal{P}_\perp$ project on $\mathscr{H}_\perp$ and $\mathscr{H}_\parallel$, respectively. We stop the protocol after the first successful detection in $\mathscr{H}_\perp$. 

\begin{figure}[t]
\centering
\includegraphics[width=\linewidth]{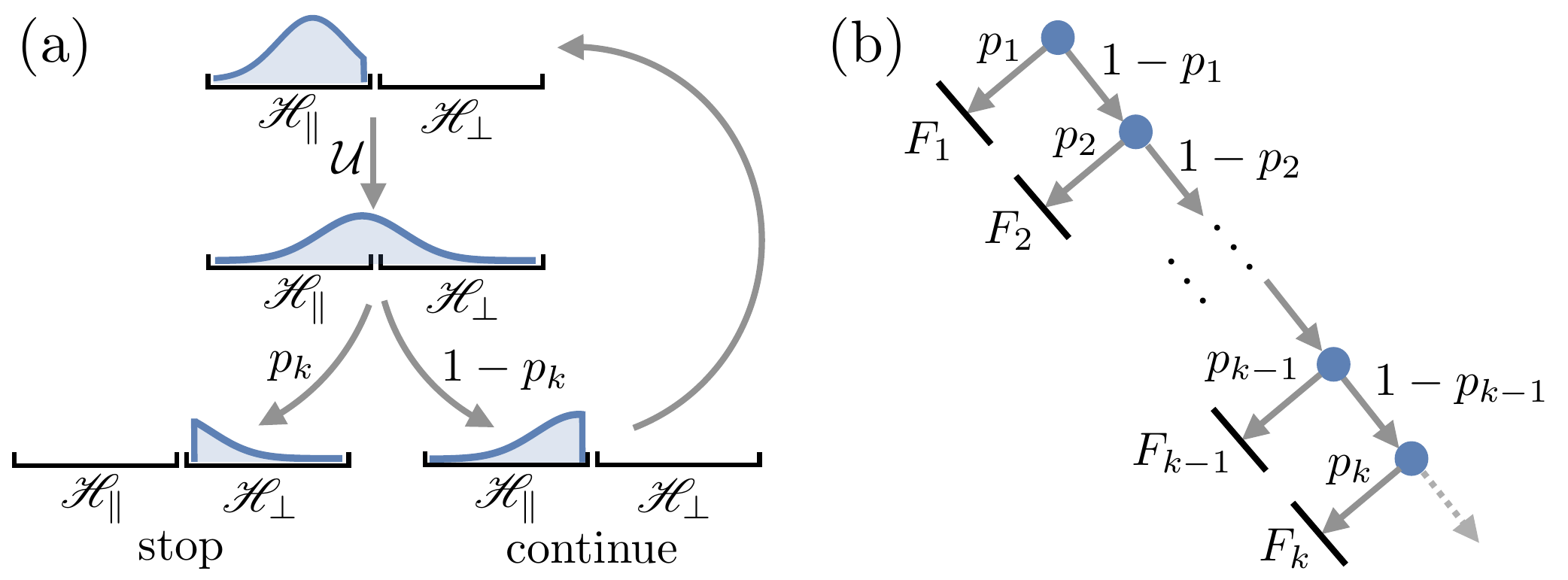}
\caption{Stroboscopic measurement. (a) A coherently evolving state (blue envelope) is repeatedly measured after equidistant time steps by a projection on the subspace $\mathscr{H}_\perp$, until successful detection. (b) The probability space is illustrated by a tree diagram, with blue points illustrating binary (i.e., yes or no) measurements.}
\label{fig:evolve}
\end{figure}

At the first measurement, the evolved state $\mathcal{U}\rho \mathcal{U}^\dagger$, with $\mathcal{U}=\exp(-\im \mathcal{H} \tau/\hbar)$, results in a successful detection with probability $p_1=\tr{\mathcal{P}_\perp \mathcal{U} \rho \mathcal{U}^\dagger \mathcal{P}_\perp}$, causing the protocol to stop. In the case of an unsuccessful detection, the state has vanishing support on $\mathscr{H}_\perp$. It reduces to $\rho_1=\mathcal{P}_\parallel \mathcal{U} \rho \mathcal{U}^\dagger \mathcal{P}_\parallel/(1-p_1)$ and continues evolving to $\mathcal{U}\rho_1\mathcal{U}^\dagger$ before the next measurement. By continuing in this way, illustrated in Fig.~\ref{fig:evolve}(a), the $k$th measurement yields a successful detection with probability
\begin{align}\label{eq:pk}
p_k=\frac{\tr{  \mathcal{P}_\perp \mathcal{U} (\mathcal{P}_\parallel \mathcal{U})^{k-1} \rho (\mathcal{U}^\dagger \mathcal{P}_\parallel )^{k-1} \mathcal{U}^\dagger \mathcal{P}_\perp  }}{(1-p_1) \cdots (1-p_{k-1})}.
\end{align}
Hence, as evident from the tree diagram in Fig.~\ref{fig:evolve}(b), the probability for the first successful detection at the $k$th measurement, called \textit{first-detection probability}, is $F_k=(1-p_1)\cdots (1-p_{k-1})p_k$. Using Eq.~\eqref{eq:pk} and the operator $\mathcal{T}_k=\mathcal{P}_\perp \mathcal{U} (\mathcal{P}_\parallel \mathcal{U})^{k-1}$, it reads
\begin{align}\label{eq:Fk}
F_k=\tr{\mathcal{T}_k^\dagger \mathcal{T}_k\rho }.
\end{align}
With this, the \textit{total detection probability} after $k$ measurements is
\begin{align}\label{eq:Dn}
D_k=\sum_{j=1}^k F_j 
\end{align}
and the \textit{survival probability}, i.e., the probability for no successful detection after $k$ measurements, becomes
\begin{align}\label{eq:Sn}
S_k=1-D_k.
\end{align}
Since the stroboscopic measurement is performed with equidistant sampling times $\tau$, the first detection probabilities~\eqref{eq:Fk} allow us to define the \textit{first-detection time} (sometimes called \cite{Friedman-QW-2017,Liu-QW-2020} \textit{mean first detection or passage time}) as the expectation value of the detection time with respect to the probability distribution $\{F_1/D_k,\dots, F_k/D_k\}$, in the limit of a large number $k$ of measurements,
\begin{align}\label{eq:tf}
\braket{t_\mathrm{f}} = \lim_{k\rightarrow \infty} \frac{1}{D_k} \sum_{j=1}^k j\tau F_j.
\end{align}
Hence, the stroboscopic measurement protocol allows us to properly define the time after which we expect the first detection in the detection subspace $\mathscr{H}_\perp$.

\subsection{Survival operator}\label{sec:SurvOp}

The probabilities $F_k$, $D_k$, and $S_k$ are key for the first-detection-time statistics in stroboscopic measurements. They are further related through Eqs.~\eqref{eq:Dn} and~\eqref{eq:Sn} by
\begin{align}\label{eq:FkSD}
F_k=D_k-D_{k-1}=S_{k-1}-S_k
\end{align}
and thus serve equally well to obtain the full statistics of the first-detection times~\eqref{eq:tf}. Under this perspective, we can now focus on rewriting one of them, the survival probability $S_k$, in a more compact and intuitive form. To this end, let us introduce the block matrix representation 
\begin{align}\label{eq:decompA}
\mathcal{A}=\begin{pmatrix}
\mathcal{A}_{\parallel\parallel}& \mathcal{A}_{\parallel\perp}\\
\mathcal{A}_{\perp\parallel}& \mathcal{A}_{\perp\perp}
\end{pmatrix}
\end{align}
of a general linear operator $\mathcal{A}$ on $\mathscr{H}$, where the first (second) column and row corresponds to the subspace $\mathscr{H}_\parallel$ ($\mathscr{H}_\perp$). Accordingly, the projectors on these subspaces are $\mathcal{P}_\parallel=\mathrm{diag}(\unit,0)$ and $\mathcal{P}_\perp=\mathrm{diag}(0,\unit)$ and the initial density matrix, which has vanishing support on the detection subspace $\mathscr{H}_\perp$, reads $\rho=\mathrm{diag}(\rho_{\parallel\parallel},0)$. As we show in Appendix~\ref{app:Sk}, we can express the survival probability~\eqref{eq:Sn} as the expectation value of the \textit{survival operator} \cite{Dhar-QT-2015,Dhar-DQ-2015}
\begin{align}\label{eq:Survival}
\mathcal{S}_k=(\mathcal{U}_{\parallel\parallel}^k)^\dagger \mathcal{U}_{\parallel\parallel}^k
\end{align}
with respect to $\rho_{\parallel\parallel}$, that is, 
\begin{align}\label{eq:SkviaOp}
S_k=\tr{\mathcal{S}_k \rho_{\parallel\parallel}}.
\end{align}
Note that, as indicated by the decomposition~\eqref{eq:decompA}, $\mathcal{U}_{\parallel\parallel}$ is the block of the unitary evolution operator $\mathcal{U}$ in the subspace $\mathscr{H}_\parallel$. Accordingly, we can intuitively interpret $S_k=\mathrm{Tr}(\mathcal{U}_{\parallel\parallel}^k \rho_{\parallel\parallel} (\mathcal{U}_{\parallel\parallel}^k)^\dagger)$ as the probability that the first $k$ measurements find the state in $\mathscr{H}_\parallel$, hence the probability for no successful detection after $k$ measurements. 

From Eqs.~\eqref{eq:tf},~\eqref{eq:FkSD},~\eqref{eq:Survival}, and~\eqref{eq:SkviaOp}, the first-detection-time statistics can be obtained by calculating the $k$th matrix power of the complex-valued matrix $\mathcal{U}_{\parallel\parallel}$ for all $k$. In general, $\mathcal{U}_{\parallel\parallel}$ is not a normal operator and hence cannot be diagonalized by a unitary matrix. However, it can be decomposed as $\mathcal{U}_{\parallel\parallel}=\mathcal{Q}\Lambda\mathcal{Q}^{-1}$ \footnote{Note that diagonalizable complex-valued matrices form a dense subset of the set of $n\times n$ complex matrices. Hence, almost all $n\times n$ complex matrices are diagonalizable and those which are not can be approximated to arbitrary precision by a diagonalizable matrix \cite{Horn-MA-2013}.}, with the columns of $\mathcal{Q}$ corresponding to the not necessarily orthogonal eigenvectors of $\mathcal{U}_{\parallel\parallel}$ and $\Lambda=\mathrm{diag}(\lambda_1,\lambda_2,\dots)$ carrying the corresponding complex-valued eigenvalues on its diagonal. Consequently, given the diagonalization $\mathcal{U}_{\parallel\parallel}=\mathcal{Q}\Lambda\mathcal{Q}^{-1}$, the matrix power $\mathcal{U}_{\parallel\parallel}^k$ can efficiently be calculated via $\mathcal{U}_{\parallel\parallel}^k=\mathcal{Q}\Lambda^k\mathcal{Q}^{-1}$, with $\Lambda^k= \mathrm{diag}(\lambda_1^k,\lambda_2^k,\dots)$.

\subsection{Trapped subspace}\label{sec:Trap}

The eigenvalues $\lambda_j$ of $\mathcal{U}_{\parallel\parallel}$ are in general complex and satisfy $0\leq |\lambda_j| \leq 1$, where the upper bound is due to $S_k\leq 1$. From this we see that for increasing $k$, all eigenvalues $\lambda_j^k$ of $\mathcal{U}_{\parallel\parallel}^k$ for which $|\lambda_j|\neq 1$ vanish exponentially in $k$. Hence, in the limit $k\rightarrow\infty$, we are left with the subspace  of $\mathcal{U}_{\parallel\parallel}$ spanned by the not necessarily orthogonal eigenvectors associated with the eigenvalues satisfying $|\lambda_j|=1$. Let us denote this subspace by $\mathscr{H}_\mathrm{T}$ and let $\mathcal{P}_\mathrm{T}$ be a projector on $\mathscr{H}_\mathrm{T}$. As we show in Appendix~\ref{app:PT}, in the limit $k\rightarrow\infty$ we find that the survival operator $\lim_{k\rightarrow \infty} \mathcal{S}_k=\mathcal{S}_\infty$ projects on the subspace $\mathscr{H}_\mathrm{T}$, i.e.,
\begin{align}\label{eq:Sinfty}
\mathcal{S}_\infty= \mathcal{P}_\mathrm{T}.
\end{align}
Accordingly, the survival probability $S_\infty = \tr{\mathcal{P}_\mathrm{T} \rho_{\parallel\parallel}}$ is equivalent to the fraction of the initial state $\rho_{\parallel\parallel}$ living on $\mathscr{H}_\mathrm{T}$ and we therefore call $\mathscr{H}_\mathrm{T}$ the \textit{trapped subspace} (note that in the literature \cite{Thiel-US-2020,Thiel-DS-2020} it is sometimes called \textit{dark subspace}). That is, in the limit $k\rightarrow\infty$ there is a nonvanishing survival probability $S_\infty$ if and only if the initial state $\rho_{\parallel\parallel}$ has support on the trapped subspace $\mathscr{H}_\mathrm{T}$ (see also \cite{Thiel-US-2020,Thiel-DS-2020}).

\subsection{Divergence of the first detection time}\label{sec:divergence}
We now continue with studying the convergence behavior of the first-detection time~\eqref{eq:tf}. First note that if $S_\infty=1$, the entire initial state remains in the trapped subspace and, from Eq.~\eqref{eq:tf}, $\braket{t_\mathrm{f}}$ is undefined. In the following we exclude this trivial case and consider $S_\infty<1$. As detailed in Appendix~\ref{app:FDT}, we first rewrite $\braket{t_\mathrm{f}}$ from Eq.~\eqref{eq:tf} as
\begin{align}\label{eq:tf02}
\frac{\braket{t_\mathrm{f}}}{\tau} =1 + \sum_{k=1}^\infty \frac{S_k-S_\infty}{1-S_\infty}.
\end{align}
The convergence properties can then be investigated via the ratio test \cite{Hildebrandt-Ana-2006} of the series in~\eqref{eq:tf02}. To this end, we consider the ratio 
\begin{align}\label{eq:ratio}
\abs{\frac{S_{k+1} - S_\infty}{S_k  - S_\infty}} \leq 1,
\end{align}
with the upper bound following from $S_{k+1} \leq S_k$. Since $\lim_{k\rightarrow\infty}S_k-S_\infty=0$, we naively expect $\braket{t_\mathrm{f}}$ from~\eqref{eq:tf02} to converge in the limit of a large number of measurements, $k\rightarrow \infty$. However, given a fixed but possibly large number of measurements $k$, it can happen that there is an additional limit in the sampling time, $\tau\rightarrow \tau_\mathrm{res}$, in which the ratio~\eqref{eq:ratio} becomes arbitrarily close to unity, hence indicating a divergent series in~\eqref{eq:tf02}. Accordingly, the first-detection time~\eqref{eq:tf02} is then characterized by a competition between the limits $\tau\rightarrow \tau_\mathrm{res}$ and $k\rightarrow \infty$, which (as discussed below) can lead to a divergent (rather than convergent) behavior of $\braket{t_\mathrm{f}}$.

Such competition arises if there is at least one continuous, differentiable eigenvalue $\lambda_j(\tau)$ of $\mathcal{U}_{\parallel\parallel}$ (note that we explicitly state the dependence of $\lambda_j$ on the sampling time $\tau$), for which $\lim_{\delta\tau \rightarrow 0}|\lambda_j(\tau_\mathrm{res}\pm\delta\tau)|=1$, and $|\lambda_j(\tau_\mathrm{res} \pm \delta\tau)|<1$ for nonvanishing but small $\delta\tau$. We call these eigenvalues $\lambda_j(\tau)$ and sampling times $\tau_\mathrm{res}$  \textit{resonant eigenvalues} and \textit{resonant sampling times}, respectively. Then $\lim_{k \rightarrow \infty} |\lambda_j^k(\tau_\mathrm{res}\pm\delta\tau)|=0$ for nonvanishing $\delta\tau$ and $\lim_{\delta\tau \rightarrow 0} |\lambda_j^k(\tau_\mathrm{res}\pm\delta\tau)|=1$ for finite $k$.

Now note that the survival operator is Hermitian [see Eq.~\eqref{eq:Survival}] and, as we show in Appendix~\ref{app:Seigendecomp}, it can be written as $\mathcal{S}_k=\mathcal{P}_\mathrm{T}+\mathcal{M}_k$, where $\mathcal{P}_\mathrm{T}$ and $\mathcal{M}_k$ have orthogonal support, and $\lim_{k \rightarrow \infty} \mathcal{M}_k=0$. Further recall that $\mathcal{P}_\mathrm{T}$ [see Eq.~\eqref{eq:Sinfty}] projects on the trapped subspace $\mathscr{H}_\mathrm{T}$, which is spanned by the eigenvectors of $\mathcal{U}_{\parallel\parallel}$ associated with eigenvalues $|\lambda_j(\tau)|=1$. Hence, for $\delta\tau=0$ the eigenvalue $\lambda_j(\tau_\mathrm{res}\pm\delta\tau)$ is associated with $\mathcal{P}_\mathrm{T}$ and for $|\delta\tau|>0$ with $\mathcal{M}_k$. Therefore, $S_k$, and hence $\braket{t_\mathrm{f}}$, can be discontinuous at $\tau_\mathrm{res}$. We further conclude [recall that $\lambda_j(\tau)$ is continuous and differentiable] that the survival operator $\mathcal{S}_k$ must have an eigenvalue, which, for small $\delta\tau$, can be well approximated by $c_m^{(k)}\approx [1-\gamma (\delta\tau)^2]^k$, where $\gamma >0$. As we discuss in Appendix~\ref{app:divergence}, if the initial state $\rho_{\parallel\parallel}$ has nonvanishing support by the corresponding eigenvector $\ket{c_m^{(k)}}$ of $\mathcal{S}_k$, i.e., $\bra{c_m^{(k)}}\rho_{\parallel\parallel}\ket{c_m^{(k)}}\neq 0$, the first-detection time diverges for small $\delta\tau$ as
\begin{align}\label{eq:tfdivergence}
\frac{\braket{t_\mathrm{f}}}{\tau} \approx A + \frac{B}{ (\delta\tau)^2},
\end{align} 
with $A,B >0$ approximately constant for small $\delta\tau$ [cf. Eqs.~(47) and~(52) in \cite{Liu-QW-2020}].

\subsection{Degenerate subspace}\label{sec:degenerate}

Next let us see how the Hamiltonian's spectral structure can lead to the appearance of resonant eigenvalues. To this end, consider the eigendecomposition $\mathcal{U}=\sum_{j} e^{-\im E_j \tau} \ketbra{E_j}{E_j}$ of the unitary evolution operator, with $\ket{E_j}$ the eigenvector associated with the eigenenergy $E_j$. Let us suppose nondegenerate eigenenergies $E_j$ and $\mathcal{P}_\parallel \ket{E_j} \neq \ket{E_j}$ [note that, if $\mathcal{P}_\parallel \ket{E_j} = \ket{E_j}$, then $\ket{E_j}$ is an eigenvector of $\mathrm{diag}(\mathcal{U}_{\parallel\parallel},0)$ whose associated eigenvalue $\lambda_j=e^{-\im E_j \tau}$ has unit modulus for any sampling time $\tau$. Degenerate eigenenergies $E_j$ can likewise lead to eigenvalues $\lambda_j$ with unit modulus for any sampling time $\tau$. However, these eigenvalues do not qualify as resonant eigenvalues as defined above]. Further suppose that there is a sampling time $\tau_\Omega$ and a corresponding set $\Omega$ of eigenenergies such that the eigenvalues $e^{-\im E_j \tau_\Omega}$ of 
$\mathcal{U}$ are degenerate for all $E_j \in\Omega$, i.e., \cite{Friedman-QR-2016,Friedman-QW-2017,Yin-LF-2019,Liu-QW-2020,Thiel-DS-2020}
\begin{align}\label{eq:taudeg}
 \forall E_j,E_k \in\Omega \ \exists \ m\in\mathbb{Z}\ : \ (E_j-E_k)\tau_\Omega = 2\pi m.
\end{align}
We then call $\tau_\Omega$ the \textit{degenerate sampling time} and define the corresponding \textit{degenerate subspace} $\mathscr{H}_\Omega \subseteq \mathscr{H}$ spanned by the eigenvectors $\{\ket{E_j}\}_{E_j \in \Omega}$, with $\mathcal{P}_\Omega=\sum_{E_j \in \Omega}\ketbra{E_j}{E_j}$ projecting on $\mathscr{H}_\Omega$. As we show in detail in Appendix~\ref{app:degenerate}, any vector $\ket{\lambda_j} \in \mathscr{H}$, which lives in the intersection of $\mathscr{H}_\Omega$ and $\mathscr{H}_\parallel$, i.e., $\mathcal{P}_\Omega\ket{\lambda_j} =\mathcal{P}_\parallel\ket{\lambda_j} =\ket{\lambda_j}$, constitutes an eigenvector of $\mathrm{diag}(\mathcal{U}_{\parallel\parallel},0)$ with associated eigenvalue $e^{-\im E_\Omega\tau_\Omega}$. Since $\ket{\lambda_j} \in \mathscr{H}$ has vanishing support on $\mathscr{H}_\perp$, i.e. $\mathcal{P}_\perp \ket{\lambda_j}=0$, we can drop the subspace $\mathscr{H}_\perp$ such that $\ket{\lambda_j}\in \mathscr{H}_\parallel$ becomes an eigenvector of $\mathcal{U}_{\parallel\parallel}$ whose associated eigenvalue $e^{-\im E_\Omega \tau_\Omega}$ has unit modulus and hence constitutes a resonant eigenvalue (see Appendix~\ref{app:degenerate} for details). As a result, we showed that a degenerate sampling time can cause a resonant sampling time.

\section{Many partially distinguishable particles}
\subsection{General formalism}\label{sec:MPGenForm}

\begin{figure}[t]
\centering
\includegraphics[width=\linewidth]{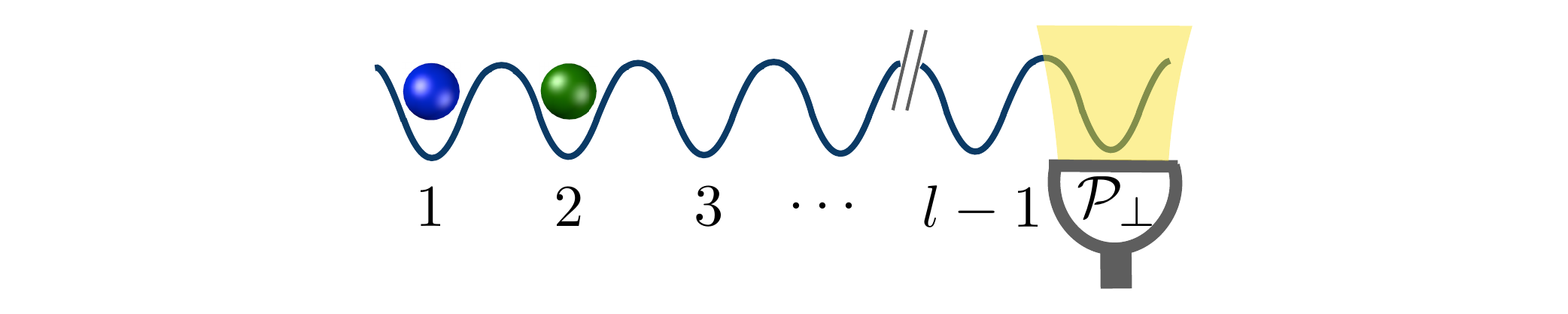}
\caption{Example of many-particle transport on a lattice. Here $N$ partially distinguishable particles (shown for $N=2$) are prepared in the first $N$ sites of a linear lattice. As the particles evolve due to tunneling dynamics, a stroboscopic projective measurement on the last lattice site $l$ is performed, with $\mathcal{P}_\perp$ the corresponding projector on the detection subspace.}
\label{fig:lattice} 
\end{figure}

Let us now apply the above formalism of stroboscopic measurements to the quantum transport of $N$ identical particles evolving across a network composed of $l$ sites and investigate how (partial) particle distinguishability affects the first-detection-time statistics via many-particle interference. As a paradigmatic setting, we consider the particles to be initially prepared on the first $N$ sites \footnote{Note that the more the particles initially bunch, the less there can be many-particle interference \cite{Dittel-AI-2019,Dittel-WP-2021}. Hence, we choose one particle per site to allow for strong interference effects.} and a detection on site $l>N$, also called the target site. Note that we impose no restrictions on the particles' hopping dynamics and on the position dependence of their mutual interaction; we merely suppose the dynamics to be governed by a general many-body unitary evolution matrix $\mathcal{U}$. As an example, which is numerically investigated in Sec.~\ref{sec:numerics} below, one can think of a linear lattice with nearest-neighbor coupling as illustrated in Fig.~\ref{fig:lattice}.

The Hilbert space $\mathscr{H}=\mathscr{H}_\parallel \oplus \mathscr{H}_\perp=\mathscr{H}_{1\mathrm{p}}^{\otimes N}$ is spanned by $N$-fold tensor products $\ket{\mathcal{E}_1}\otimes\dots \otimes\ket{\mathcal{E}_N}\equiv \ket{\mathcal{E}_1\cdots \mathcal{E}_N}$ of the single-particle basis states $\{\ket{\mathcal{E}_j} \}_{\mathcal{E}_j=1}^l$ of $\mathscr{H}_{1\mathrm{p}}$, with $\ket{\mathcal{E}_j}$ describing a single particle on site $\mathcal{E}_j$. For the initial state of one particle in each of the first $N$ sites, we have $\vec{\mathcal{E}}=(\mathcal{E}_1\cdots \mathcal{E}_N)=(1\cdots N)$. Note that the case of bunched initial states can straightforwardly be described following Refs.~\cite{Dittel-AI-2019,Dittel-WP-2021}.

Particle distinguishability is accounted for by additional internal degrees of freedom on an $N$-particle Hilbert space $\mathscr{H}_\mathrm{int}$. We make no restrictions on the particles' nonsymmetrized internal states, described by a density operator $\sigma$ on $\mathscr{H}_\mathrm{int}$ (e.g., a pure product state gives $\sigma=\ket{\vec{\phi}}\bra{\vec{\phi}}$, with $\ket{\vec{\phi}}=\ket{\phi_1}\otimes \cdots \otimes\ket{\phi_N}$). 

The initial state $\rho$ on $\mathscr{H}$ of $N$ partially distinguishable bosons (fermions) is then obtained by (anti)symmetrizing $\ketbra{\vec{\mathcal{E}}}{\vec{\mathcal{E}}} \otimes \sigma$ on $\mathscr{H}\otimes \mathscr{H}_\mathrm{int}$ with respect to all elements $\pi \in \mathrm{S}_N$ of the symmetric group $\mathrm{S}_N$ of $N$ elements and tracing over the particles' internal degrees of freedom. Using the shorthand notation $\Pi_\pi\ket{\vec{\mathcal{E}}}=\ket{\vec{\mathcal{E}}_\pi}=\ket{\mathcal{E}_{\pi(1)}\cdots \mathcal{E}_{\pi(N)}}$, with permutations $\pi\in\mathrm{S}_N$, and $\Pi_\pi$ the corresponding permutation operator, this yields \cite{Dittel-AI-2019,Minke-CF-2021,Dittel-WP-2021}
\begin{align}\label{eq:rho}
\rho= \sum_{\pi,\pi' \in \mathrm{S_N}} \rho_{\pi,\pi'} \ketbra{\vec{\mathcal{E}}_\pi}{\vec{\mathcal{E}}_{\pi'}},
\end{align}
with
\begin{align}
\rho_{\pi,\pi'} = (-1)^{\pi\pi'}_\mathrm{B(F)}\frac{1}{N!} \tr{\Pi_\pi \sigma \Pi^\dagger_{\pi'}}.
\end{align}
Here $(-1)^{\pi\pi'}_\mathrm{B}=1$ for bosons, $(-1)^{\pi\pi'}_\mathrm{F}=\sgn(\pi\pi')$ for fermions, and $\Pi_\pi$ permutes $\ket{\vec{\phi}} \in \mathscr{H}_\mathrm{int}$ similarly to $\ket{\vec{\mathcal{E}}} \in \mathscr{H}$. For perfectly indistinguishable particles, $\rho$ is pure and has maximal many-body coherences, i.e., off-diagonal elements, $\rho=\ketbra{\psi_\mathrm{B(F)}}{\psi_\mathrm{B(F)}}$, with $\ket{\psi_\mathrm{B(F)}}=(1/\sqrt{N!}) \sum_{\pi\in\mathrm{S}_N} (-1)^\pi_\mathrm{B(F)}\ket{\vec{\mathcal{E}}_\pi}$ the usual pure state of $N$ indistinguishable bosons (fermions). On the other hand, for fully distinguishable particles we have $\rho_{\pi,\pi'} =\delta_{\pi,\pi'}/N!$. This yields a maximally mixed state $\rho=(1/N!) \sum_{\pi \in \mathrm{S}_N} \ketbra{\vec{\mathcal{E}}_\pi}{\vec{\mathcal{E}}_\pi}$, which is fully incoherent \cite{Dittel-AI-2019,Dittel-WP-2021}. 

The coherences of $\rho$ are essential for many-particle interference in the particles' transport dynamics. Hence, they are relevant for the first-detection-time statistics in stroboscopic measurements as directly apparent from Eq.~\eqref{eq:SkviaOp}. To this end, let us quantify the many-body coherences of $\rho$ via the expectation value of the projector $\Pi_\mathrm{S}=(1/N!) \sum_{\pi \in \mathrm{S}_N} \Pi_\pi$ on the symmetric subspace \cite{Minke-CF-2021,Brunner-MB-2022},
\begin{align}\label{eq:PiS}
\braket{\Pi_\mathrm{S}}&= \tr{ \Pi_\mathrm{S} \rho}\\
&=\frac{1}{N!}\sum_{\pi,\pi'\in\mathrm{S}_N} \rho_{\pi,\pi'}.\label{eq:PiSelem}
\end{align}
It corresponds to a sum of all elements of $\rho$ and satisfies $0 \leq \braket{\Pi_\mathrm{S}} \leq 1$. The upper bound saturates for a fully symmetric state (e.g., fully indistinguishable bosons), the lower bound is reached if $\rho$ has no support on the symmetric subspace (e.g., if two or more fermions are fully indistinguishable), and fully distinguishable particles, which give rise to a fully incoherent state, result in $ \braket{\Pi_\mathrm{S}}=1/N!$.

For the stroboscopic measurement, we consider different binary projective measurements: the detection of exactly $n$ ($=n$), and at least $n$ ($\geq n$) particles at the target site. As we show in Appendix~\ref{app:MeasurementsOp}, the projectors on the corresponding detection subspaces can be expressed as
\begin{align}\label{eq:Pperpn}
\mathcal{P}^{=n}_\perp=\sum_{q=n}^N (-1)^{q-n}{q \choose n} \mathcal{P}_q
\end{align}
and
\begin{align}\label{eq:Pperpgeqn}
\mathcal{P}^{\geq n}_\perp=\sum_{q=n}^N (-1)^{q-n} \frac{n}{q} {q \choose n} \mathcal{P}_q,
\end{align}
with $\mathcal{P}_q$ a genuine $q$-particle observable given by
\begin{align}\label{eq:Pq}
\mathcal{P}_q=\sum_{\sigma \in \Sigma(L^{(q)})} \bigotimes_{\alpha=1}^N \mathcal{A}_{L_{\sigma(\alpha)}^{(q)}}.
\end{align}
Here $\mathcal{A}_1=\ketbra{l}{l}$, $\mathcal{A}_2=\unit$, $L_j^{(q)}$ is the $j$th entry of the multiset \footnote{A multiset is a generalization of a set where multiple instances of the same element are allowed.} $L^{(q)}=\{1\}^q \cup \{2\}^{N-q}$, and $\Sigma(L^{(q)}) \subseteq \mathrm{S}_N$ is the right transversal \footnote{A \textit{transversal} of a collection of sets $B_1,\dots,B_R$ is a set of $R$ elements which contains exactly one element of each set $B_1,\dots,B_R$. For $H$ a subgroup of the group $G$, the \textit{right transversal} of $H$ in $G$ is a transversal of the set of distinct right cosets of $H$ in $G$. The right coset of $H$ in $G$ with respect to $\pi \in H$ is $H\pi=\{ \xi \pi\ | \ \xi\in H \}$ \cite{Baumslag-SO-1968}.} of the Young subgroup $\mathrm{S}_q\otimes\mathrm{S}_{N-q}$ in $\mathrm{S}_N$ containing all ${N \choose q}$ permutations $\sigma$, which lead to distinctly ordered multisets. For example, for $N=2$ particles, we have the one-particle observable $\mathcal{P}_1=\ketbra{l}{l} \otimes \unit + \unit \otimes \ketbra{l}{l}$ and the two-particle observable $\mathcal{P}_2=\ketbra{l}{l} \otimes\ketbra{l}{l}$.  The projection operators on the detection subspaces can then be expressed as $\mathcal{P}^{=1}_\perp=\mathcal{P}_1-2\mathcal{P}_2=\ketbra{l}{l}\otimes(\unit-\ketbra{l}{l})+(\unit-\ketbra{l}{l})\otimes \ketbra{l}{l}$, $\mathcal{P}^{=2}_\perp=\mathcal{P}_2$, and $\mathcal{P}^{\geq 1}_\perp=\mathcal{P}_1-\mathcal{P}_2=\mathcal{P}^{=1}_\perp+\mathcal{P}^{=2}_\perp$. A comparison of Eqs.~\eqref{eq:Pperpn} and~\eqref{eq:Pperpgeqn} shows that both expressions merely differ by the fraction $n/q$ in~\eqref{eq:Pperpgeqn}. Since $\mathcal{P}_q$ is a genuine $q$-particle observable and this fraction decreases for increasing $q$, we see that the sensitivity of $\mathcal{P}^{\geq n}_\perp$ to genuine $q$-particle effects (e.g., $q$-particle interference) decreases for increasing $q$ as compared to $\mathcal{P}^{=n}_\perp$. Hence, in accordance with the typical interpretation of (anti)bunching as a manifestation of many-particle properties~\cite{Carolan-OE-2014,Shchesnovich-UG-2016}, for the measurement $\mathcal{P}^{=n}_\perp$ we expect more pronounced many-particle interference effects in the first-detection-time statistics compared to $\mathcal{P}^{\geq n}_\perp$.

The stroboscopic measurement corresponding to the detection of a fully bunched event on the target site, i.e., the detection of exactly $N$ particles with projector $\mathcal{P}^{=N}_\perp$, is particularly interesting. In this case, we find that the first-detection probability~\eqref{eq:Fk} is weighted by the fraction~\eqref{eq:PiS} of $\rho$ on the symmetric subspace. In particular, as we show in Appendix~\ref{app:FkBunched}, we find
\begin{align}\label{eq:FkBunched}
F_k=N! \braket{\Pi_\mathrm{S}} F_k^\mathrm{D},
\end{align}
with $F_k^\mathrm{D}$ the first-detection probability in the case of fully distinguishable particles.  Accordingly, the total detection probability~\eqref{eq:Dn} becomes $D_k=N! \braket{\Pi_\mathrm{S}} D_k^\mathrm{D}$ and the survival probability~\eqref{eq:Sn}
\begin{align}\label{eq:SkequalN},
S_k=1-N!\braket{\Pi_\mathrm{S}} \left(1-S_k^\mathrm{D}  \right).
\end{align}
Using this for the first-detection time $\braket{t_\mathrm{f}}$ in Eq.~\eqref{eq:tf}, we see that the factor $N!\braket{\Pi_\mathrm{S}}$ cancels such that $\braket{t_\mathrm{f}}$ is independent on the particles' distinguishability, i.e., 
\begin{align}\label{eq:tfequal}
\braket{t_\mathrm{f}}=\braket{t_\mathrm{f}^\mathrm{D}},
\end{align}
with $\braket{t_\mathrm{f}^\mathrm{D}}$ the first-detection time for fully distinguishable particles. However, note that $\braket{t_\mathrm{f}}$ is undefined if $\braket{\Pi_\mathrm{S}}=0$ since then $D_\infty=0$ [cf. Eq.~\eqref{eq:tf}]. This, for example, applies to two or more fully indistinguishable fermions, as a consequence of Pauli's exclusion principle, which prohibits the detection of indistinguishable fermions on the same site. 

\subsection{Numerical illustration}\label{sec:numerics}

We now turn to the numerical investigation of an example for the above-described stroboscopic measurements in many-particle quantum transport. To this end we consider $N=2$ noninteracting, partially distinguishable particles initially prepared in the first two sites of a linear lattice with $l>N$ sites (see Fig.~\ref{fig:lattice}). We suppose nearest-neighbor tunneling, with equal tunneling rates between neighboring sites, and ask for the first-detection time to detect at least one ($\geq 1$), exactly one ($=1$), and exactly two ($=2$) particles at the target site $l$.  

For the particles' internal state we consider a pure product state $\ket{\vec{\phi}}=\ket{\phi_1}\otimes \ket{\phi_2} \in \mathscr{H}_\text{int}$, with $\ket{\phi_j}$ the internal state of the $j$th particle such that, after (anti)symmetrization, the initial state~\eqref{eq:rho} reads
\begin{align}\label{eq:initwopartstate}
\begin{split}
\rho&=\frac{1}{2} \Big[ \ketbra{12}{12} + \ketbra{21}{21} \\
&\pm \abs{\bracket{\phi_1}{\phi_2}}^2 \left( \ketbra{12}{21} + \ketbra{21}{12}\right)  \Big],
\end{split}
\end{align}
with the upper (lower) sign corresponding to bosons (fermions). Note that the particles' indistinguishability is encoded in the off-diagonal entries of $\rho$ and quantified by the expectation value of the projector on the symmetric subspace~\eqref{eq:PiSelem}, yielding $\braket{\Pi_\mathrm{S}}=(1 \pm \abs{\bracket{\phi_1}{\phi_2}}^2 )/2$.

The linear lattice with constant tunneling rate $J$ between neighboring sites is described by the single-particle Hamiltonian
\begin{align}
\mathcal{H}_{1\mathrm{p}}=- J \sum_{\mathcal{E}=1}^{l-1} \left( \ketbra{\mathcal{E}}{\mathcal{E}+1} + \ketbra{\mathcal{E}+1}{\mathcal{E}} \right),
\end{align}
which generates the single-particle unitary evolution operator $\mathcal{U}_{1\mathrm{p}}=\exp(-\im \mathcal{H}_{1\mathrm{p}} \tau/\hbar)$. Since we consider noninteracting particles, the two-particle unitary evolution operator is obtained from the tensor power thereof, $\mathcal{U}=\mathcal{U}_{1\mathrm{p}}^{\otimes 2}$.

\begin{figure*}[t]
\centering
\includegraphics[width=\linewidth]{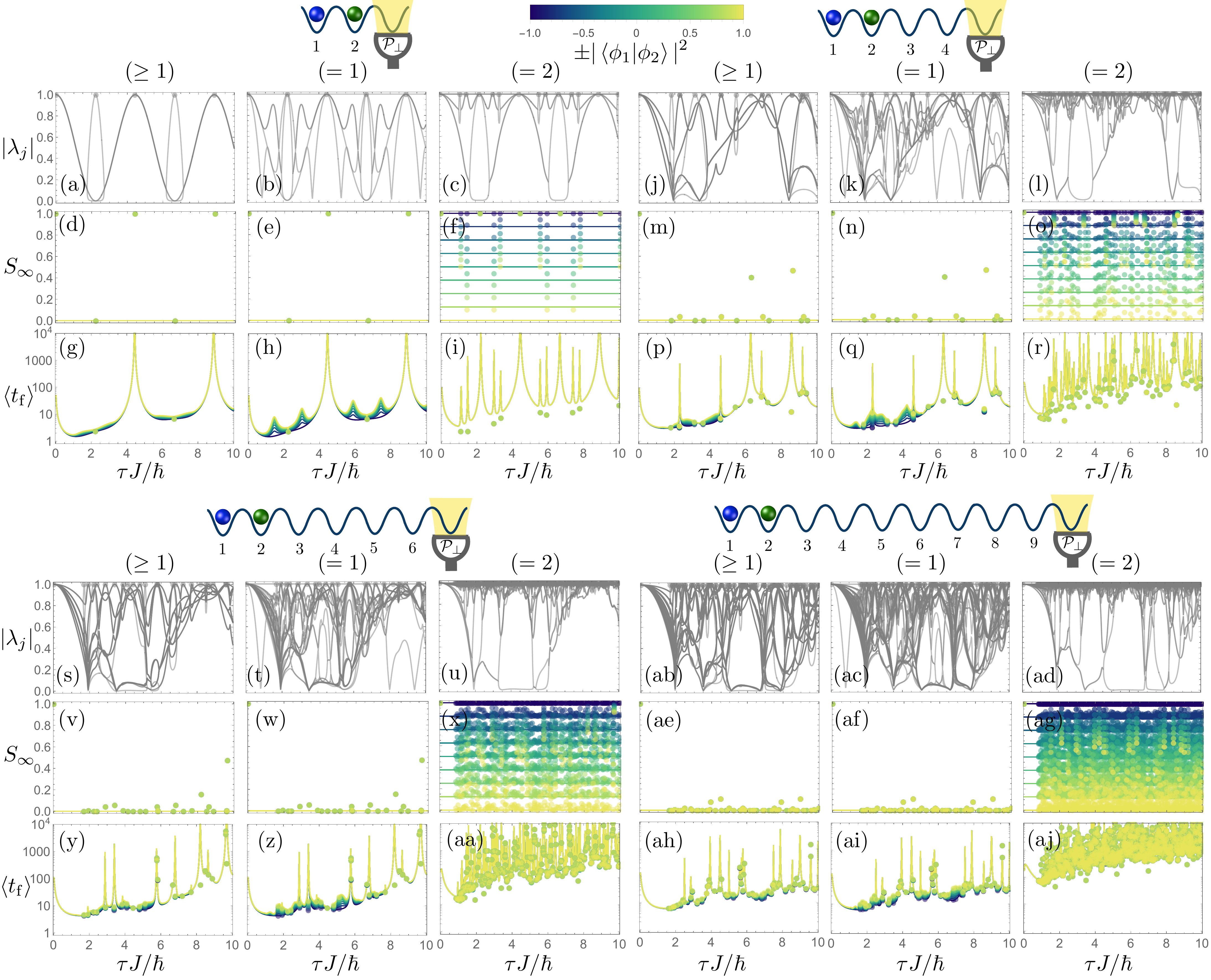}
\caption{Numerical results for the transport of two particles across a linear lattice under stroboscopic measurements. (a)-(i) For a linear lattice with nearest-neighbor tunneling rate $J$ between $l=3$ sites, the first, second, and third columns show the results for the stroboscopic measurement of at least one ($\geq 1$), exactly one ($=1$), and exactly two ($=2$) particles at the target site $l$, respectively. In (a)-(c) the moduli $\abs{\lambda_j}$ of the eigenvalues of the block matrix $\mathcal{U}_{\parallel\parallel}$ are shown as a function of the sampling time $\tau J/\hbar$. Resonant eigenvalues (which give rise to resonant sampling times) are marked by closed circles and multiple eigenvalues lying on top of each other cause the solid line to appear darker. (d)-(f) Survival probability $S_\infty$ plotted against the sampling time $\tau J/\hbar$. (g)-(i) First-detection time $\braket{t_\mathrm{f}}$ as a function of $\tau J/\hbar$. Closed circles in (d)-(i) indicate the values of $S_\infty$ and $\braket{t_\mathrm{f}}$ for the resonant sampling times and the coloring corresponds to different particle distinguishabilities $\pm |\langle \phi_1 |\phi_2 \rangle|^2$ [see Eq.~\eqref{eq:initwopartstate}], as indicated by the color bar. (j)-(aj) Results similar to (a)-(i) but for a lattice with (j)-(r) $l=5$, (s)-(aa) $l=7$, and (ab)-(aj) $l=10$ sites.}
\label{fig:3sites} 
\end{figure*}

With the two-particle unitary evolution operator $\mathcal{U}$ and the initial two-particle state $\rho$ at hand, we calculate the moduli of the eigenvalues of $\mathcal{U}_{\parallel\parallel}$, the survival probability $S_\infty$, and the first-detection time $\braket{t_\mathrm{f}}$ [using Eqs.~\eqref{eq:SkviaOp}-\eqref{eq:tf02}] for $501$ equally spaced sampling times in the interval $0 \leq \tau J/\hbar \leq 10$ as well as for all resonant sampling times, nine different particle distinguishabilities equally spaced in the interval $-1\leq \pm \abs{\bracket{\phi_1}{\phi_2}}^2 \leq 1$, and different lattice lengths $l=3,\ 5,\ 7,$ and $10$. In our calculation of the first-detection time, the sum in Eq.~\eqref{eq:tf02} is truncated if $S_k-S_\infty < 10^{-4}$ or $k>10^4$ due to limited computation time. In the vicinity of divergences of the first-detection time $\braket{t_\mathrm{f}}$, this truncation as well as the finite resolution of the sampling time may result in a peak instead of a true divergence to infinity [for example, in Fig.~\ref{fig:3sites}(p)].

Let us start by discussing the linear lattice with $l=3$ sites. In this particular case, the particles' dynamics are perfectly periodic for sampling times $\tau J/\hbar$, which are integer multiples of $2\pi/\sqrt{2}\approx 4.44$, i.e., for  $\tau J/\hbar= z 2\pi/\sqrt{2}$, with $z\in\mathbb{N}_0$, we have $\mathcal{U}=\unit$. As a result, we see a periodic behavior for the moduli $\abs{\lambda_j}$ of the eigenvalues of $\mathcal{U}_{\parallel\parallel}$ in Figs.~\ref{fig:3sites}(a)-\ref{fig:3sites}(c). Here resonant eigenvalues with associated resonant sampling times $\tau_\mathrm{res}$ are marked by closed circles. As discussed in detail in Sec.~\ref{sec:divergence}, depending on the initial state, the first-detection time  $\braket{t_\mathrm{f}}$ can diverge for these resonant sampling times. This is evident by comparing the resonant sampling times in Figs.~\ref{fig:3sites}(a)-\ref{fig:3sites}(c) with the divergences of the first-detection time in Figs.~\ref{fig:3sites}(g)-\ref{fig:3sites}(i). Note that for the measurement of at least one and exactly one particle at the target site, the divergences of $\braket{t_\mathrm{f}}$ in Figs.~\ref{fig:3sites}(g) and~\ref{fig:3sites}(h) only appear if $\tau J/\hbar$ is an integer multiple of $2\pi/\sqrt{2}$. These are the sampling times for which $\mathcal{U}=\unit$ due to the periodic dynamics, with all eigenvalues of $\mathcal{U}_{\parallel\parallel}$ satisfying $\abs{\lambda_j}=1$ [see Figs.~\ref{fig:3sites}(a) and~\ref{fig:3sites}(b)]. Accordingly, for these sampling times the initial state remains unaffected such that nothing can be detected at the target site, as indicated by the survival probability $S_\infty =1$ in Figs.~\ref{fig:3sites}(d) and~\ref{fig:3sites}(e). In the vicinity of these sampling times, the unitary evolution operator slightly deviates from the identity. As a result, the divergence of $\braket{t_\mathrm{f}}$ at these sampling times can be understood through the quantum Zeno effect \cite{Degasperis-DL-1974}. 

For the measurement of exactly two particles at the target site, there are additional divergences of $\braket{t_\mathrm{f}}$ for resonant sampling times, which are no integer multiples of $2\pi/\sqrt{2}$ [see Fig.~\ref{fig:3sites}(i)]. These divergences can be understood through the competition of the limits $\delta\tau\rightarrow 0$ and $k\rightarrow \infty$ in the vicinity of resonant eigenvalues $\lambda_j^k(\tau_\mathrm{res}\pm\delta\tau)$ (see Sec.~\ref{sec:divergence}). First consider the divergences appearing halfway between the periodic sampling times, at $\tau J/\hbar  = (z+1/2)2\pi/\sqrt{2}$, with $z\in\mathbb{N}_0$, i.e., for $\tau J/\hbar \approx 2.22, 6.66, \dots$. For these sampling times, the unitary evolution operator mirrors the initial state of the particles with respect to the second site such that one particle occupies site $2$ and one particle site $3$. This state has no support on the detection subspace for the measurement of exactly two particles at the target site. Hence, we have a unit survival probability $S_\infty= 1$ [see Fig.~\ref{fig:3sites}(f)] and the first-detection time at these sampling times is undefined [see above Eq.~\eqref{eq:tf02}]. However, for small nonvanishing $\delta\tau$ around these sampling times, the state is not perfectly mirrored between consecutive measurements. In this case, $\braket{t_\mathrm{f}}$ is well defined and must diverge in the limit $\delta\tau\rightarrow 0$ for which the survival probability $S_\infty$ becomes unity. On the other hand, at the divergences which are not exactly halfway between periodic sampling times, e.g., for $\tau J/\hbar \approx 1.11, 1.48,\dots$, the survival probability $S_\infty$ in Fig.~\ref{fig:3sites}(f) becomes unity only for perfectly indistinguishable fermions (which is due to Pauli's exclusion principle). As a result, except for indistinguishable fermions, the first-detection time $\braket{t_\mathrm{f}}$ takes a finite value at these resonant sampling times, indicated by the closed circles in Fig.~\ref{fig:3sites}(i).

Next let us consider the effect of particle distinguishability for the survival probability $S_\infty$ and the first-detection time $\braket{t_\mathrm{f}}$ shown in Figs.~\ref{fig:3sites}(d)-\ref{fig:3sites}(i). First, consider the measurement of exactly two particles. In this case, we see that the survival probability shown in Fig.~\ref{fig:3sites}(f) follows Eq.~\eqref{eq:SkequalN}, which for the two-particle state~\eqref{eq:initwopartstate} becomes $S_k=S_k^\mathrm{D} \mp \abs{\bracket{\phi_1}{\phi_2}}^2(1-S_k^\mathrm{D})$, with the upper (lower) sign corresponding to bosons (fermions). Furthermore, as predicted by Eq.~\eqref{eq:tfequal}, Fig.~\ref{fig:3sites}(i) reveals that the first-detection time $\braket{t_\mathrm{f}}$ appears independently of the particles' indistinguishability. For the measurement of at least one and exactly one particle at the target site, Figs.~\ref{fig:3sites}(d) and~\ref{fig:3sites}(e) show a vanishing survival probability $S_\infty=0$, independently of the particles' distinguishability, except at integer multiples of the period $2\pi/\sqrt{2}$. However, between the divergences, the first-detection times in Figs.~\ref{fig:3sites}(g) and~\ref{fig:3sites}(h) show a clear dependence on the particles' indistinguishability: The more symmetric the two-particle state, i.e., the larger $\pm \abs{\bracket{\phi_1}{\phi_2}}^2$ in~\eqref{eq:initwopartstate}, the larger the first detection time $\braket{t_\mathrm{f}}$ \footnote{Note that an increasing first-detection time for increasing symmetry of the many-body state is not a generally valid trend. Instead, it is an artifact of the chosen lattice Hamiltonian. For example, for a one-dimensional linear lattice with three sites and periodic boundary conditions, i.e., for a ring, and for the detection of exactly one particle, we find sampling times for which the first-detection time increases with decreasing symmetry of the two-body state \cite{Neubrand-FD-2020}.}. As expected from our considerations below Eq.~\eqref{eq:Pq}, this effect appears more pronounced for the measurement of exactly one particle compared to the measurement of at least one particle.

Next we increase the lattice length to $l=5$ sites. The numerical results are shown in Figs.~\ref{fig:3sites}(j)-\ref{fig:3sites}(r). In this case, there are no perfectly periodic dynamics such that the survival probability yields $S_\infty=1$ only if $\tau J/\hbar=0$ or (due to Pauli's principle) if we consider the measurement of exactly two particles in the case of perfectly indistinguishable fermions [see Figs.~\ref{fig:3sites}(m)-\ref{fig:3sites}(o)]. However, as indicated in Figs.~\ref{fig:3sites}(j)-\ref{fig:3sites}(l), there is an increasing number of resonant eigenvalues, which give rise to resonant sampling times for which divergences appear in the first-detection time if the initial state has nonvanishing support by the corresponding eigenvector (see Sec.~\ref{sec:divergence}). Indeed, Figs.~\ref{fig:3sites}(p)-\ref{fig:3sites}(r) show multiple divergences of $\braket{t_\mathrm{f}}$ together with the values of the first-detection time for the exact resonant sampling times. The behavior of the survival probability $S_\infty$ and the first-detection time $\braket{t_\mathrm{f}}$ as a function of the particles' distinguishability is similar to the behavior on the lattice with $l=3$ sites. If we further increase the lattice length to $l=7$ and $10$ sites [see Figs.~\ref{fig:3sites}(s)-\ref{fig:3sites}(aj)], we again find an increasing number of resonant eigenvalues $|\lambda_j|$ and an increasing number of divergences of the first-detection time. Interestingly, the general behavior of $|\lambda_j|$, $S_\infty$, and $\braket{t_\mathrm{f}}$ appears similar to the case with lattice length $l=5$, up to a rescaling of the time axis. Note that the purpose of Figs.~\ref{fig:3sites}(j)-\ref{fig:3sites}(aj) is not that the reader can decipher all details, but to see that the general structure of the first-detection-time statistics remains similar for an increasing lattice length.

\section{Summary and conclusion}\label{sec:Conclusion}
Stroboscopic measurements provide a natural approach to investigate first-detection times in quantum evolutions, a problem of longstanding interest \cite{Allock-TA-1969}. While recent investigations \cite{Dhar-QT-2015,Dhar-DQ-2015,Friedman-QR-2016,Friedman-QW-2017,Thiel-FD-2018,Yin-LF-2019,Liu-QW-2020,Thiel-US-2020,Thiel-DS-2020,Kessler-FD-2021} mainly focused on pure single-particle states, we provided here a description in the density operator formalism, which allowed us to apply the concept of stroboscopic measurements to the realm of many partially distinguishable and possibly interacting particles, where particle indistinguishability enters through the coherences of the corresponding many-body density operator. Now, on the many-body level, binary projective-valued measurements which are sensitive to the particle number are conceivable. We focused here on the detection of exactly $n$ and at least $n$ particles at a single target site and showed that the former is more sensitive to genuine many-body effects such as many-body interference, a result in accordance with the widespread wisdom that (anti)bunching is a typical manifestation of genuine many-body indistinguishability \cite{Mayer-CS-2011,Carolan-OE-2014,Shchesnovich-UG-2016}. Moreover, for the detection of all particles at the target, we found that for an increasing symmetry of the many-body state, the increasing probability for perfect bunching balances with the decreasing probability for no successful detection such that, by normalization, the first-detection time appears independently of the particles' distinguishability. Other binary measurements include, e.g., the detection of coincidences on two different target sites, which we investigated numerically (not shown here) but did not observe strong differences to the numerical data presented here. Our results constitute only a first step towards a rigorous understanding of the first-detection-time statistics on the many-body level and leave many open questions for future research: What is the role of particle interactions versus particle indistinguishability? How does the first return time \cite{Gruenbaum-RD-2013,Yin-LF-2019,Liu-QW-2020} behave? What is the effect of random sampling times \cite{Kessler-FD-2021}? What happens in the limit of large particle numbers? Which kind of binary measurements are experimentally feasible?

\begin{acknowledgments}
The authors thank Jonathan Brugger, Dominik Lentrodt, and Moritz Richter for fruitful discussions. C.D. acknowledges the Georg H. Endress Foundation for support and the Freiburg Institute for Advanced Studies for a FRIAS Junior Fellowship.
\end{acknowledgments}

\begin{appendix}
\section{Proof of Eq.~(\ref{eq:SkviaOp})}\label{app:Sk}
In order to prove Eq.~\eqref{eq:SkviaOp}, we use Eqs.~\eqref{eq:Dn} and~\eqref{eq:Sn} to write $S_k=1-\sum_{j=1}^k F_j$. Using $F_k$ from Eq.~\eqref{eq:Fk} and $\rho=\mathcal{P}_\parallel \rho \mathcal{P}_\parallel$, we have
\begin{align}\label{eq:FkAppendix}
\begin{split}
F_k&=\tr{\mathcal{T}^\dagger_k\mathcal{T}_k \rho}\\
&=\tr{\mathcal{P}_\parallel\mathcal{T}^\dagger_k\mathcal{T}_k \mathcal{P}_\parallel \rho}.
\end{split}
\end{align}
With $\mathcal{T}_k=\mathcal{P}_\perp \mathcal{U} (\mathcal{P}_\parallel \mathcal{U})^{k-1}$ [see above Eq.~\eqref{eq:Fk}] and using that $\mathcal{P}_\perp$ and $\mathcal{P}_\parallel$ are projectors, we have
\begin{align}
\begin{split}
\mathcal{P}_\parallel\mathcal{T}^\dagger_k\mathcal{T}_k \mathcal{P}_\parallel&=\mathcal{P}_\parallel(\mathcal{U}^\dagger \mathcal{P}_\parallel )^{k-1} \mathcal{U}^\dagger \mathcal{P}_\perp \mathcal{U}  (\mathcal{P}_\parallel \mathcal{U})^{k-1} \mathcal{P}_\parallel\\
&=(\mathcal{P}_\parallel\mathcal{U}^\dagger \mathcal{P}_\parallel )^{k-1} \mathcal{U}^\dagger \mathcal{P}_\perp \mathcal{U}  (\mathcal{P}_\parallel \mathcal{U} \mathcal{P}_\parallel)^{k-1}.
\end{split}
\end{align}
Inserting $\mathcal{P}_\perp=\unit - \mathcal{P}_\parallel$ yields
\begin{align}
&\phantom{=} \mathcal{P}_\parallel\mathcal{T}^\dagger_k\mathcal{T}_k \mathcal{P}_\parallel \nonumber \\
&=(\mathcal{P}_\parallel\mathcal{U}^\dagger \mathcal{P}_\parallel )^{k-1} \mathcal{U}^\dagger (\unit - \mathcal{P}_\parallel) \mathcal{U}  (\mathcal{P}_\parallel \mathcal{U} \mathcal{P}_\parallel)^{k-1}\\
&=(\mathcal{P}_\parallel\mathcal{U}^\dagger \mathcal{P}_\parallel )^{k-1}  (\mathcal{P}_\parallel \mathcal{U} \mathcal{P}_\parallel)^{k-1} - (\mathcal{P}_\parallel\mathcal{U}^\dagger \mathcal{P}_\parallel )^k (\mathcal{P}_\parallel \mathcal{U} \mathcal{P}_\parallel)^k. \nonumber 
\end{align}
By plugging this into Eq.~\eqref{eq:FkAppendix} and using $\rho=\mathrm{diag}(\rho_{\parallel\parallel},0)$, $\mathcal{P}_\parallel \mathcal{U} \mathcal{P}_\parallel=\mathrm{diag}(\mathcal{U}_{\parallel\parallel},0)$, and $\mathcal{P}_\parallel \mathcal{U}^\dagger \mathcal{P}_\parallel=\mathrm{diag}(\mathcal{U}_{\parallel\parallel}^\dagger,0)$ we find
\begin{align}
\begin{split}
F_k&=\tr{\left[ (\mathcal{U}_{\parallel\parallel}^\dagger)^{k-1}\mathcal{U}_{\parallel\parallel}^{k-1} - (\mathcal{U}_{\parallel\parallel}^\dagger)^k\mathcal{U}_{\parallel\parallel}^k\right] \rho_{\parallel\parallel}}\\
&=\tr{\left[ \mathcal{S}_{k-1} - \mathcal{S}_k\right] \rho_{\parallel\parallel}}.
\end{split}
\end{align}
In the last step we used $(\mathcal{U}_{\parallel\parallel}^\dagger)^k=(\mathcal{U}_{\parallel\parallel}^k)^\dagger$ and the definition of $\mathcal{S}_k$ from Eq.~\eqref{eq:Survival}. With $S_k=1-\sum_{j=1}^k F_j$ we finally arrive at
\begin{align}
\begin{split}
S_k&=1-\sum_{j=1}^k \tr{ \left[ \mathcal{S}_{j-1} - \mathcal{S}_j\right] \rho_{\parallel\parallel}}\\
&=1-\tr{\left[ \unit - \mathcal{S}_k\right] \rho_{\parallel\parallel}}\\
&=\tr{\mathcal{S}_k \rho_{\parallel\parallel}},
\end{split}
\end{align}
which coincides with Eq.~\eqref{eq:SkviaOp}.

\section{Proof of Eq.~(\ref{eq:Sinfty})}\label{app:PT}
Since the survival operator $\mathcal{S}_\infty$ is Hermitian [see Eq.~\eqref{eq:Survival}], with associated survival probability $0\leq S_\infty = \tr{\mathcal{S}_\infty \rho_{\parallel\parallel}}\leq 1$, it has an eigendecomposition 
\begin{align}\label{eq:SinftyEigendecomp}
\mathcal{S}_\infty=\sum_m c_m \ketbra{c_m}{c_m},
\end{align}
with positive eigenvalues $0\leq c_m \leq 1$ and $\{\ket{c_m}\}_m$ an orthonormal eigenbasis of $\mathscr{H}_\parallel$. As discussed above Eq.~\eqref{eq:Sinfty} in the main text, the set of normalized, not necessarily orthogonal eigenvectors $\{\ket{\lambda_j}\}_j$ of $\mathcal{U}_{\parallel\parallel}$ also forms a basis of $\mathscr{H}_\parallel$. These eigenvectors can be uniquely written as $\ket{\lambda_j}=\sum_m \mathcal{D}_{m,j} \ket{c_m}$, with $\sum_m |\mathcal{D}_{m,j}|^2=1$. This defines the invertible basis transformation matrix $\mathcal{D}$ with elements $\mathcal{D}_{m,j}=\bracket{c_m}{\lambda_j}=\bra{c_m}\mathcal{D}\ket{c_j}$, acting as
\begin{align}
\begin{split}
\ket{\lambda_j} &= \mathcal{D}\ket{c_j},\\
\ket{c_j}&= \mathcal{D}^{-1} \ket{\lambda_j}.
\end{split}
\end{align}
Together with Eq.~\eqref{eq:SinftyEigendecomp} we then find
\begin{align}\label{eq:AppSinft}
\begin{split}
\bra{\lambda_j}\mathcal{S}_\infty \ket{\lambda_j}&= \sum_m c_m \abs{\bracket{c_m}{\lambda_j}}^2\\
&=\sum_m c_m \abs{ \mathcal{D}_{m,j}}^2.
\end{split}
\end{align}
On the other hand, using Eq.~\eqref{eq:Survival} and $\mathcal{U}_{\parallel\parallel}^k\ket{\lambda_j}=\lambda_j^k\ket{\lambda_j}$, we have
\begin{align}
\begin{split}
\bra{\lambda_j}\mathcal{S}_\infty \ket{\lambda_j}&=\lim_{k\rightarrow\infty} \abs{\lambda_j}^{2k}\\
&=\begin{cases} 1 & \text{for } \abs{\lambda_j}=1\\ 0& \text{otherwise} \end{cases}
\end{split}
\end{align}
such that, with Eq.~\eqref{eq:AppSinft},
\begin{align}\label{eq:Delements}
\sum_m c_m \abs{ \mathcal{D}_{m,j}}^2 = \begin{cases} 1 & \text{for } \abs{\lambda_j}=1\\ 0& \text{otherwise}. \end{cases}
\end{align}
Now suppose that $\dim(\mathscr{H}_\parallel)=d$ and that there are $n$ eigenvalues satisfying $|\lambda_j|=1$, which we collect by the index set $J=\{j\,:\,|\lambda_j|=1\}$. Consider the set of columns $\{ \mathcal{D}_{m,j} \}_{0\leq m \leq d, j\in J}$: Since $\mathcal{D}$ is invertible, this set of columns has at least $n$ distinct rows, which contain a nonvanishing element $\mathcal{D}_{m,j}\neq 0$. Accordingly, from Eq.~\eqref{eq:Delements}, $\sum_m|\mathcal{D}_{m,j}|^2=1$, and $c_m\leq 1$, $\mathcal{S}_\infty$ must have at least $n$ eigenvalues $c_m=1$. We collect these eigenvalues by the index set $M=\{m\,:\,c_m=1\}$. Now consider the set of columns $\{ \mathcal{D}_{m,j} \}_{0\leq m \leq d, j\notin J}$: Similar to before, since $\mathcal{D}$ is invertible, this set of columns has at least $d-n$ distinct rows, which contain a nonvanishing element $\mathcal{D}_{m,j}\neq 0$. However, since $|\lambda_j| \leq 1$ such that $\lim_{k\rightarrow \infty} |\lambda_j|^{2k}=0$, for $j\notin J$, from Eq.~\eqref{eq:Delements}, there must be at least $d-n$ eigenvalues $c_m=0$. Hence, we find that there are at least $n$ eigenvalues $c_m=1$ and at least $d-n$ eigenvalues $c_m=0$. Given that $\mathcal{S}_\infty$ has exactly $d$ eigenvalues, we conclude that it has exactly $n$ eigenvalues $c_m=1$ and $d-n$ eigenvalues $c_m=0$. This implies that $\{\ket{\lambda_j}\}_{j\in J}$ and $\{\ket{c_m}\}_{m\in M}$ are isomorphic, with the survival operator from Eq.~\eqref{eq:SinftyEigendecomp} reading
\begin{align}\label{eq:B6}
\begin{split}
\mathcal{S}_\infty &= \sum_{m\in M} \ketbra{c_m}{c_m}\\
&= \mathcal{P}_\mathrm{T},
\end{split}
\end{align}
where $\mathcal{P}_\mathrm{T}$ is the projector on the trapped subspace $\mathscr{H}_\mathrm{T}$ spanned by the not necessarily orthogonal eigenvectors $\{ \ket{\lambda_j}\}_{j \in J}$ associated with the eigenvalues of $\mathcal{U}_{\parallel\parallel}$ satisfying $|\lambda_j|=1$.

\section{Proof of Eq.~(\ref{eq:tf02})}\label{app:FDT}
We prove Eq.~\eqref{eq:tf02} by starting from Eq.~\eqref{eq:tf} and using Eq.~\eqref{eq:FkSD},
\begin{align}
\begin{split}
\braket{t_\mathrm{f}} &=\lim_{k\rightarrow \infty} \frac{1}{D_k} \sum_{j=1}^k j\tau F_j\\
&=\lim_{k\rightarrow \infty} \frac{\tau}{D_k} \sum_{j=1}^k j(D_j-D_{j-1})\\
&=\lim_{k\rightarrow \infty} \frac{\tau}{D_k} \left[ \sum_{j=1}^k j D_j - \sum_{j'=0}^{k-1} (j'+1) D_{j'} \right].
\end{split}
\end{align}
With $D_0=0$, this expression simplifies to
\begin{align}
\begin{split}
\braket{t_\mathrm{f}}=&\lim_{k\rightarrow \infty} \frac{\tau}{D_k}  \left[ k D_k - \sum_{j=1}^{k-1} D_j\right]\\
&= \lim_{k\rightarrow \infty} \tau \left[k+1 - \sum_{j=1}^{k}\frac{D_j}{D_k} \right] \\
&=\lim_{k\rightarrow \infty}  \tau \left[ 1 + \sum_{j=1}^k \left( 1 - \frac{D_j}{D_k} \right)\right].
\end{split}
\end{align}
Finally, using Eq.~\eqref{eq:Sn}, we arrive at
\begin{align}
\braket{t_\mathrm{f}}=\tau \left[ 1 + \sum_{j=1}^\infty \frac{S_j-S_\infty}{1-S_\infty}\right],
\end{align}
which coincides with Eq.~\eqref{eq:tf02}.

\section{Eigendecomposition of $\mathcal{S}_k$}\label{app:Seigendecomp}

In the following we show that $\mathcal{S}_k$ can be written as $\mathcal{S}_k=\mathcal{P}_\mathrm{T}+\mathcal{M}_k$, with $\mathcal{P}_\mathrm{T}$ and $\mathcal{M}_k$ having orthogonal support, and $\lim_{k\rightarrow\infty} \mathcal{M}_k=0$. First, let us collect all eigenstates $\ket{\lambda_j}$ of $\mathcal{U}_{\parallel \parallel}$ whose corresponding eigenvalues have unit modulus. To this end, we define the index set $J=\{ j\ :\ |\lambda_j|=1\}$. Accordingly, by the definition of $\mathcal{S}_k$ from Eq.~\eqref{eq:Survival}, we have, for all $j\in J$,
\begin{align}
\bra{\lambda_j} \mathcal{S}_k \ket{\lambda_j}=1.
\end{align}
Since the survival probability~\eqref{eq:SkviaOp} satisfies $\braket{\mathcal{S}_k}=S_k \leq 1$, this shows that for all $j\in J$ the ket $\ket{\lambda_j}$ must also be an eigenstate of $\mathcal{S}_k$ with corresponding eigenvalue $1$, i.e., the set of states $\{ \ket{\lambda_j}\}_{j \in J}$ spans a degenerate subspace of $\mathcal{S}_k$ to the eigenvalue $1$ (note that, for finite $k$, the survival operator $\mathcal{S}_k$ might have further eigenvalues equal to $1$ whose corresponding eigenvectors have vanishing support on this subspace). As discussed above Eq.~\eqref{eq:Sinfty} and explicitly proven in Appendix~\ref{app:PT} [see below Eq.~\eqref{eq:B6}], this degenerate subspace is the trapped subspace. Accordingly, the eigendecomposition of $\mathcal{S}_k$ can be written as
\begin{align}\label{eqapp:eigendecomp}
\mathcal{S}_k = \mathcal{P}_\mathrm{T} + \sum_m c_m^{(k)} \ketbra{c_m^{(k)}}{c_m^{(k)}},
\end{align}
where $\mathcal{P}_\mathrm{T}$ is the projector on the trapped subspace and $c_m^{(k)} \leq 1$ are eigenvalues of $\mathcal{S}_k$ with corresponding eigenvector $\ket{c_m^{(k)}}$. By defining $\mathcal{M}_k=\sum_m c_m^{(k)} \ketbra{c_m^{(k)}}{c_m^{(k)}}$, we then see that $\mathcal{S}_k=\mathcal{P}_\mathrm{T} +\mathcal{M}_k$, with $\mathcal{P}_\mathrm{T}$ and $\mathcal{M}_k$ having orthogonal support and, from Eq.~\eqref{eq:Sinfty}, $\lim_{k\rightarrow\infty}\mathcal{M}_k=0$.

\section{Proof of Eq.~(\ref{eq:tfdivergence})}\label{app:divergence}
Let us prove the divergence behavior of the first-detection time stated in Eq.~\eqref{eq:tfdivergence} by considering the expression of $\braket{t_\mathrm{f}}$ from Eq.~\eqref{eq:tf02},
\begin{align}\label{eq:tfappss}
\frac{\braket{t_\mathrm{f}}}{\tau} =1 + \sum_{k=1}^\infty \frac{S_k-S_\infty}{1-S_\infty}.
\end{align}
Using $S_k=\tr{\mathcal{S}_k \rho_{\parallel\parallel}}$ and the eigendecomposition $\mathcal{S}_k=\mathcal{P}_\mathrm{T} + \sum_m c_m^{(k)} \ketbra{c_m^{(k)}}{c_m^{(k)}}$ of the survival operator [see Eq.~\eqref{eqapp:eigendecomp}], where $\mathcal{P}_\mathrm{T}=\mathcal{S}_\infty$ [see Eq.~\eqref{eq:Sinfty}] projects on the trapped subspace $\mathscr{H}_\mathrm{T}$ and $c_m^{(k)}$ are eigenvalues of $\mathcal{S}_k$ with corresponding eigenvectors $\ket{c_m^{(k)}}$, the numerator in the series in Eq.~\eqref{eq:tfappss} can be written as
\begin{align}
S_k-S_\infty=\tr{ \sum_m c_m^{(k)} \ketbra{c_m^{(k)}}{c_m^{(k)}} \rho_{\parallel\parallel}}.
\end{align}
Let us now suppose that there is at least one resonant eigenvalue such that there must be at least one eigenvalue $c_m^{(k)}$, which is well approximated by $[1-\gamma  (\delta\tau)^2]^k$ for small $\delta\tau$ (see the main text for details). Next recall that all other eigenvalues $c_m^{(k)}$ must vanish much faster for large $k$. Hence, for large $k$, we have 
\begin{align}\label{eq:SkSinftapp}
S_k-S_\infty \approx [1-\gamma (\delta\tau)^2]^k \tilde{B},
\end{align}
with $\tilde{B}$ approximately constant for small $\delta\tau$ and nonvanishing if and only if $\rho_{\parallel\parallel}$ has support on at least one of the eigenvectors $\ket{c_m^{(k)}}$ corresponding to the eigenvalues which are well approximated by $[1-\gamma  (\delta\tau)^2]^k$. Given that the approximation in~\eqref{eq:SkSinftapp} only holds for large $k$, we can introduce $\tilde{A} \in \mathbb{R}$ (which is also approximately constant for small $\delta\tau$) and rewrite Eq.~\eqref{eq:tfappss} as
\begin{align}
\frac{\braket{t_\mathrm{f}}}{\tau} \approx \tilde{A} + \frac{\tilde{B}}{1-S_\infty} \sum_{k=1}^\infty [1-\gamma  (\delta\tau)^2]^k.
\end{align}
Using the geometric series $\sum_{k=0}^\infty a^k=1/(1-a)$ and merging all constants finally leads to 
\begin{align}
\frac{\braket{t_\mathrm{f}}}{\tau} \approx A+  \frac{B}{(\delta\tau)^2},
\end{align}
where $B=\tilde{B}/\gamma(1-S_\infty)$ and $A=\tilde{A}-\tilde{B}/(1-S_\infty)$.

\section{Resonant eigenvalues}\label{app:degenerate}
For degenerate sampling time $\tau_\Omega$, the unitary $\mathcal{U}$ reads [see Eq.~\eqref{eq:taudeg}]
\begin{align}
\mathcal{U}=e^{-\im E_\Omega \tau_\Omega} \mathcal{P}_\Omega + \sum_{E_j \neq \Omega} e^{-\im E_j \tau_\Omega} \ketbra{E_j}{E_j},
\end{align}
where $\mathcal{P}_\Omega=\sum_{E_j \in \Omega}\ketbra{E_j}{E_j}$. Further recall that $\mathcal{P}_\parallel \ket{E_j} \neq \ket{E_j}$ for all $E_j$ [see above Eq.~\eqref{eq:taudeg}]. Now consider any vector $\ket{\lambda_j} \in \mathscr{H}$, which lives on the intersection of $\mathscr{H}_\parallel$ and $\mathscr{H}_\Omega$, satisfying $\mathcal{P}_\parallel\ket{\lambda_j}=\mathcal{P}_\Omega\ket{\lambda_j}=\ket{\lambda_j}$. Using $\mathcal{P}_\Omega\ket{\lambda_j}=\ket{\lambda_j}$, we find $\mathcal{U}\ket{\lambda_j}=e^{-\im E_\Omega\tau_\Omega}\ket{\lambda_j}$. Together with $\mathcal{P}_\parallel\ket{\lambda_j}=\ket{\lambda_j}$, this yields
\begin{align}
\begin{split}
\mathrm{diag}(\mathcal{U}_{\parallel\parallel},0)\ket{\lambda_j}&=\mathcal{P}_\parallel\mathcal{U}\mathcal{P}_\parallel\ket{\lambda_j} \\
&=\mathcal{P}_\parallel\mathcal{U}\ket{\lambda_j}\\
&=e^{-\im E_\Omega\tau_\Omega}\mathcal{P}_\parallel\ket{\lambda_j}\\
&=e^{-\im E_\Omega\tau_\Omega}\ket{\lambda_j}.
\end{split}
\end{align}
Accordingly, $\ket{\lambda_j} \in \mathscr{H}$ is an eigenvector of $\mathrm{diag}(\mathcal{U}_{\parallel\parallel},0)$ with eigenvalue $e^{-\im E_\Omega\tau_\Omega}$. Given that $\mathcal{P}_\perp\ket{\lambda_j}=0$, we can drop the subspace $\mathscr{H}_\perp$ such that $\ket{\lambda_j} \in \mathscr{H}_\parallel$ becomes an eigenvector of $\mathcal{U}_{\parallel\parallel}$ with eigenvalue $\lambda_j=e^{-\im E_\Omega\tau_\Omega}$. Given that $\lim_{\delta\tau\rightarrow 0}|\lambda_j(\tau_\Omega\pm\delta\tau)|=1$ and  $|\lambda_j(\tau_\Omega\pm\delta\tau)|<1$ for small but nonvanishing $\delta\tau$ (the latter is due to the fact that the degenerate subspace only forms for the sampling time $\tau_\Omega$), we can identify $\lambda_j$ as a resonant eigenvalue, with resonant sampling time $\tau_\mathrm{res}=\tau_\Omega$. From this we see that a degenerate sampling time can cause a resonant sampling time.

\section{Proof of Eqs.~(\ref{eq:Pperpn}) and~(\ref{eq:Pperpgeqn}) }\label{app:MeasurementsOp}
In the following, we prove the expression for the projector $\mathcal{P}_\perp^{=n}$ and $\mathcal{P}_\perp^{\geq n}$ from Eqs.~\eqref{eq:Pperpn} and~\eqref{eq:Pperpgeqn}, respectively. We start with $\mathcal{P}_\perp^{=n}$. To this end, let us first consider the case $n=1$, i.e., the detection of exactly one particle at the target site $l$. In this case, we have
\begin{align}\label{eq:appPg1} 
\mathcal{P}_\perp^{=1}=\sum_{\alpha=1}^N\mathcal{B}_2^{\otimes \alpha -1}\otimes \mathcal{B}_1 \otimes \mathcal{B}_2^{\otimes N-\alpha},
\end{align}
where $\mathcal{B}_1=\ketbra{l}{l}$ and $\mathcal{B}_2=\unit-\ketbra{l}{l}$. Note that $\mathcal{P}_\perp^{=1}$ is symmetric under particle exchange, as required for an operator acting on many identical particles. Using the multiset of $N$ indices $L^{(1)}=\{1\}^1\cup\{2\}^{N-1}=\{1,2,\dots,2\}$, Eq.~\eqref{eq:appPg1} can be rewritten as
\begin{align}\label{eq:appPg1O}
\mathcal{P}_\perp^{=1}=\sum_{\sigma \in \Sigma(L^{(1)})}^N \bigotimes_{\alpha=1}^N \mathcal{B}_{L_{\sigma(\alpha)}^{(1)}},
\end{align}
where $L_j^{(1)}$ is the $j$th element of $L^{(1)}$ and $\Sigma(L^{(1)})=\{(1~2),(1~3),\dots,(1~N)\} \subseteq \mathrm{S}_N$ is the right transversal of $\mathrm{S}_1\otimes \mathrm{S}_{N-1}$ in $\mathrm{S}_N$ containing all permutations $\sigma$, which lead to distinctly ordered multisets, with permutations $\sigma$ provided in cycle notation. For a general number $n$, the projector $\mathcal{P}^{=n}_\perp$ can be written similarly to in~\eqref{eq:appPg1O}. In this case, the multiset generalizes to $L^{(n)}=\{1\}^n\cup\{2\}^{N-n}$, with $\Sigma(L^{(n)})$ the right transversal of the Young subgroup $\mathrm{S}_n \otimes \mathrm{S}_{N-n}$ in $\mathrm{S}_N$, containing ${N \choose n}$ permutations $\sigma$, which lead to distinct multisets. Accordingly, we have
\begin{align}
\mathcal{P}_\perp^{=n}=\sum_{\sigma \in \Sigma(L^{(n)})}^N \bigotimes_{\alpha=1}^N \mathcal{B}_{L_{\sigma(\alpha)}^{(n)}}.
\end{align}
Next we multiply out all operators $\mathcal{B}_2=\unit-\ketbra{l}{l}$. To this end, let us define $\mathcal{A}_1=\ketbra{l}{l}$, $\mathcal{A}_{-1}=-\ketbra{l}{l}=-\mathcal{A}_1$, and $\mathcal{A}_2=\unit$ such that $\mathcal{B}_1=\mathcal{A}_1$ and $\mathcal{B}_2=\mathcal{A}_2+\mathcal{A}_{-1}$. Furthermore, let us introduce the multiset of $N$ indices $\mathcal{L}^{(n,q)}=\{-1\}^{q-n} \cup \{1\}^n \cup \{2\}^{N-q}$. After a moment of thought, we see that 
\begin{align}
\mathcal{P}_\perp^{=n}=\sum_{q=n}^N\sum_{\sigma \in \Sigma(\mathcal{L}^{(n,q)})}^N \bigotimes_{\alpha=1}^N \mathcal{A}_{\mathcal{L}_{\sigma(\alpha)}^{(n,q)}}.
\end{align}
Since $\mathcal{A}_1$ and $\mathcal{A}_{-1}$ only differ by their sign, the tensor product can be rewritten in terms of the multiset $L^{(q)}$, $\bigotimes_{\alpha=1}^N \mathcal{A}_{\mathcal{L}_{\sigma(\alpha)}^{(n,q)}}=(-1)^{q-n} \bigotimes_{\alpha=1}^N \mathcal{A}_{L_{\sigma(\alpha)}^{(q)}}$. If we also want to convert the sum over all $\sigma \in \Sigma(\mathcal{L}^{(n,q)})$ into a sum over all $\sigma \in \Sigma(L^{(q)})$, we additionally have to introduce the normalization factor $|\mathcal{L}^{(n,q)}|/|L^{(q)}|$, with $|\mathcal{L}^{(n,q)}|=N!/(N-q)!(q-n)! n!$ the cardinality of $\mathcal{L}^{(n,q)}$ and $|L^{(q)}|={N \choose q}$ the cardinality of $L^{(q)}$. Altogether, this yields
\begin{align}
\begin{split}
\mathcal{P}_\perp^{=n}&=\sum_{q=n}^N(-1)^{q-n} \frac{N!}{(N-q)!(q-n)! n! {N \choose q}} \\
&\times \sum_{\sigma \in \Sigma(L^{(q)})}^N \bigotimes_{\alpha=1}^N \mathcal{A}_{L_{\sigma(\alpha)}^{(q)}}\\
&=\sum_{q=n}^N(-1)^{q-n} {q \choose n} \sum_{\sigma \in \Sigma(L^{(q)})}^N \bigotimes_{\alpha=1}^N \mathcal{A}_{L_{\sigma(\alpha)}^{(q)}}.
\end{split}
\end{align}
Using the definition of the genuine $q$-particle observable $\mathcal{P}_q$ from Eq.~\eqref{eq:Pq}, we arrive at
\begin{align}\label{eq:appPgnproved}
\mathcal{P}^{=n}_\perp=\sum_{q=n}^N (-1)^{q-n}{q \choose n} \mathcal{P}_q,
\end{align}
which coincides with the sought-after relation from Eq.~\eqref{eq:Pperpn}.

Next we prove Eq.~\eqref{eq:Pperpgeqn}. Using 
\begin{align}
\mathcal{P}^{\geq n}_\perp = \sum_{n'=n}^N \mathcal{P}^{=n'}_\perp
\end{align}
and inserting $\mathcal{P}^{=n'}_\perp$ from Eq.~\eqref{eq:appPgnproved} yields
\begin{align}
\mathcal{P}^{\geq n}_\perp = \sum_{n'=n}^N\sum_{q=n'}^N (-1)^{q-n'}{q \choose n'} \mathcal{P}_q.
\end{align}
We now use that ${q \choose n'}=0$ for $q<n'$ such that the second sum can be started at $q=n$ instead of $q=n'$,
\begin{align}
\mathcal{P}^{\geq n}_\perp &= \sum_{n'=n}^N\sum_{q=n}^N (-1)^{q-n'}{q \choose n'} \mathcal{P}_q.
\end{align}
Given that the two sums are now independent of each other, we can exchange their order and then, similarly to before, lower the upper limit of the second sum from $n'=N$ to $n'=q$, 
\begin{align}
\begin{split}
\mathcal{P}^{\geq n}_\perp &= \sum_{q=n}^N\sum_{n'=n}^N (-1)^{q-n'}{q \choose n'} \mathcal{P}_q\\
&=\sum_{q=n}^N\sum_{n'=n}^q (-1)^{q-n'}{q \choose n'} \mathcal{P}_q.
\end{split}
\end{align}
Finally, performing the sum over $n'$ results in the expression for $\mathcal{P}^{\geq n}_\perp$ from Eq.~\eqref{eq:Pperpgeqn},
\begin{align}
\mathcal{P}^{\geq n}_\perp=\sum_{q=n}^N (-1)^{q-n} \frac{n}{q} {q \choose n} \mathcal{P}_q.
\end{align}

\section{Proof of Eq.~(\ref{eq:FkBunched})}\label{app:FkBunched}
Let us prove Eq.~\eqref{eq:FkBunched} by starting with $F_k=\mathrm{Tr}(\mathcal{T}_k^\dagger \mathcal{T}_k\rho )$ from Eq.~\eqref{eq:Fk}, where $\mathcal{T}_k=\mathcal{P}_\perp \mathcal{U} (\mathcal{P}_\parallel \mathcal{U})^{k-1}$. First, we write $\mathcal{T}_k=\mathcal{P}_\perp \mathcal{J}_k$, with  $\mathcal{J}_k= \mathcal{U} (\mathcal{P}_\parallel \mathcal{U})^{k-1}$, such that
\begin{align}\label{eq:appFkBun1}
F_k=\tr{\mathcal{J}_k^\dagger \mathcal{P}_\perp \mathcal{J}_k \rho}.
\end{align}
Given that we consider the detection of exactly $N$ particles on the target site $l$, the corresponding projector~\eqref{eq:Pperpn} onto the detection subspace reads $\mathcal{P}_\perp^{=N}=\ketbra{L}{L}$, with $\ket{L}=\ket{l}\otimes \dots \otimes \ket{l}$. Using this in Eq.~\eqref{eq:appFkBun1}, we get
\begin{align}
F_k=\bra{L} \mathcal{J}_k \rho \mathcal{J}_k^\dagger \ket{L}.
\end{align}
By plugging in $\rho$ from Eq.~\eqref{eq:rho}, this yields
\begin{align}\label{eq:appFkBun2}
F_k= \sum_{\pi,\pi' \in \mathrm{S_N}} \rho_{\pi,\pi'} \bra{L}\mathcal{J}_k \ket{\vec{\mathcal{E}}_\pi} \bra{\vec{\mathcal{E}}_{\pi'}}\mathcal{J}_k^\dagger \ket{L}.
\end{align}
Next we use that for all $\pi\in\mathrm{S}_N$ we have $\ket{L}=\Pi_\pi\ket{L}$ and $\mathcal{J}_k\Pi_\pi= \Pi_\pi\mathcal{J}_k$. The latter is due to $\mathcal{J}_k$ being an operator acting on many identical particles. Hence, we have 
\begin{align}
\begin{split}
\bra{L}\mathcal{J}_k \ket{\vec{\mathcal{E}}_\pi}&=\bra{L}\mathcal{J}_k \Pi_\pi\ket{\vec{\mathcal{E}}}\\
&=\bra{L}\mathcal{J}_k\ket{\vec{\mathcal{E}}}
\end{split}
\end{align}
such that Eq.~\eqref{eq:appFkBun2} becomes
\begin{align}\label{eq:appFkBun3}
\begin{split}
F_k&= \sum_{\pi,\pi' \in \mathrm{S_N}} \rho_{\pi,\pi'} \bra{L}\mathcal{J}_k \ket{\vec{\mathcal{E}}} \bra{\vec{\mathcal{E}}}\mathcal{J}_k^\dagger \ket{L}\\
&=\abs{\bra{L}\mathcal{J}_k \ket{\vec{\mathcal{E}}}}^2\sum_{\pi,\pi' \in \mathrm{S_N}} \rho_{\pi,\pi'} .
\end{split}
\end{align}
Using that the density matrix elements corresponding to fully distinguishable particles satisfy $\rho_{\pi,\pi'}=\delta_{\pi,\pi'}/N!$, Eq.~\eqref{eq:appFkBun3} leads us to $F_k^\mathrm{D}=|\bra{L}\mathcal{J}_k \ket{\vec{\mathcal{E}}}|^2$. Hence, together with Eq.~\eqref{eq:PiSelem}, we arrive at
\begin{align}
F_k= N! \braket{\Pi_\mathrm{S}} F_k^\mathrm{D},
\end{align}
which coincides with Eq.~\eqref{eq:FkBunched}.

\end{appendix}


\begin{thebibliography}{44}%
\makeatletter
\providecommand \@ifxundefined [1]{%
 \@ifx{#1\undefined}
}%
\providecommand \@ifnum [1]{%
 \ifnum #1\expandafter \@firstoftwo
 \else \expandafter \@secondoftwo
 \fi
}%
\providecommand \@ifx [1]{%
 \ifx #1\expandafter \@firstoftwo
 \else \expandafter \@secondoftwo
 \fi
}%
\providecommand \natexlab [1]{#1}%
\providecommand \enquote  [1]{``#1''}%
\providecommand \bibnamefont  [1]{#1}%
\providecommand \bibfnamefont [1]{#1}%
\providecommand \citenamefont [1]{#1}%
\providecommand \href@noop [0]{\@secondoftwo}%
\providecommand \href [0]{\begingroup \@sanitize@url \@href}%
\providecommand \@href[1]{\@@startlink{#1}\@@href}%
\providecommand \@@href[1]{\endgroup#1\@@endlink}%
\providecommand \@sanitize@url [0]{\catcode `\\12\catcode `\$12\catcode
  `\&12\catcode `\#12\catcode `\^12\catcode `\_12\catcode `\%12\relax}%
\providecommand \@@startlink[1]{}%
\providecommand \@@endlink[0]{}%
\providecommand \url  [0]{\begingroup\@sanitize@url \@url }%
\providecommand \@url [1]{\endgroup\@href {#1}{\urlprefix }}%
\providecommand \urlprefix  [0]{URL }%
\providecommand \Eprint [0]{\href }%
\providecommand \doibase [0]{https://doi.org/}%
\providecommand \selectlanguage [0]{\@gobble}%
\providecommand \bibinfo  [0]{\@secondoftwo}%
\providecommand \bibfield  [0]{\@secondoftwo}%
\providecommand \translation [1]{[#1]}%
\providecommand \BibitemOpen [0]{}%
\providecommand \bibitemStop [0]{}%
\providecommand \bibitemNoStop [0]{.\EOS\space}%
\providecommand \EOS [0]{\spacefactor3000\relax}%
\providecommand \BibitemShut  [1]{\csname bibitem#1\endcsname}%
\let\auto@bib@innerbib\@empty
\bibitem [{\citenamefont {Farhi}\ and\ \citenamefont
  {Gutmann}(1998)}]{Farhi-QC-1998}%
  \BibitemOpen
  \bibfield  {author} {\bibinfo {author} {\bibfnamefont {E.}~\bibnamefont
  {Farhi}}\ and\ \bibinfo {author} {\bibfnamefont {S.}~\bibnamefont
  {Gutmann}},\ }\bibfield  {title} {\bibinfo {title} {Quantum computation and
  decision trees},\ }\href {https://doi.org/10.1103/PhysRevA.58.915} {\bibfield
   {journal} {\bibinfo  {journal} {Physical Review A}\ }\textbf {\bibinfo
  {volume} {58}},\ \bibinfo {pages} {915} (\bibinfo {year} {1998})}\BibitemShut
  {NoStop}%
\bibitem [{\citenamefont {Ambainis}\ \emph {et~al.}(2001)\citenamefont
  {Ambainis}, \citenamefont {Bach}, \citenamefont {Nayak}, \citenamefont
  {Vishwanath},\ and\ \citenamefont {Watrous}}]{Ambainis-OD-2001}%
  \BibitemOpen
  \bibfield  {author} {\bibinfo {author} {\bibfnamefont {A.}~\bibnamefont
  {Ambainis}}, \bibinfo {author} {\bibfnamefont {E.}~\bibnamefont {Bach}},
  \bibinfo {author} {\bibfnamefont {A.}~\bibnamefont {Nayak}}, \bibinfo
  {author} {\bibfnamefont {A.}~\bibnamefont {Vishwanath}},\ and\ \bibinfo
  {author} {\bibfnamefont {J.}~\bibnamefont {Watrous}},\ }\bibfield  {title}
  {\bibinfo {title} {One-dimensional quantum walks},\ }in\ \href
  {https://doi.org/10.1145/380752.380757} {\emph {\bibinfo {booktitle}
  {Proceedings of the Thirty-Third Annual ACM Symposium on Theory of
  Computing}}},\ \bibinfo {series and number} {STOC '01}\ (\bibinfo
  {publisher} {Association for Computing Machinery},\ \bibinfo {address} {New
  York, NY, USA},\ \bibinfo {year} {2001})\ p.\ \bibinfo {pages}
  {37–49}\BibitemShut {NoStop}%
\bibitem [{\citenamefont {Varbanov}\ \emph {et~al.}(2008)\citenamefont
  {Varbanov}, \citenamefont {Krovi},\ and\ \citenamefont
  {Brun}}]{Varbanov-HT-2008}%
  \BibitemOpen
  \bibfield  {author} {\bibinfo {author} {\bibfnamefont {M.}~\bibnamefont
  {Varbanov}}, \bibinfo {author} {\bibfnamefont {H.}~\bibnamefont {Krovi}},\
  and\ \bibinfo {author} {\bibfnamefont {T.~A.}\ \bibnamefont {Brun}},\
  }\bibfield  {title} {\bibinfo {title} {Hitting time for the continuous
  quantum walk},\ }\href {https://doi.org/10.1103/PhysRevA.78.022324}
  {\bibfield  {journal} {\bibinfo  {journal} {Physical Review A}\ }\textbf
  {\bibinfo {volume} {78}},\ \bibinfo {pages} {022324} (\bibinfo {year}
  {2008})}\BibitemShut {NoStop}%
\bibitem [{\citenamefont {Gr{\"u}nbaum}\ \emph {et~al.}(2013)\citenamefont
  {Gr{\"u}nbaum}, \citenamefont {Vel{\'a}zquez}, \citenamefont {Werner},\ and\
  \citenamefont {Werner}}]{Gruenbaum-RD-2013}%
  \BibitemOpen
  \bibfield  {author} {\bibinfo {author} {\bibfnamefont {F.~A.}\ \bibnamefont
  {Gr{\"u}nbaum}}, \bibinfo {author} {\bibfnamefont {L.}~\bibnamefont
  {Vel{\'a}zquez}}, \bibinfo {author} {\bibfnamefont {A.~H.}\ \bibnamefont
  {Werner}},\ and\ \bibinfo {author} {\bibfnamefont {R.~F.}\ \bibnamefont
  {Werner}},\ }\bibfield  {title} {\bibinfo {title} {Recurrence for discrete
  time unitary evolutions},\ }\href {https://doi.org/10.1007/s00220-012-1645-2}
  {\bibfield  {journal} {\bibinfo  {journal} {Communications in Mathematical
  Physics}\ }\textbf {\bibinfo {volume} {320}},\ \bibinfo {pages} {543}
  (\bibinfo {year} {2013})}\BibitemShut {NoStop}%
\bibitem [{\citenamefont {Degasperis}\ \emph {et~al.}(1974)\citenamefont
  {Degasperis}, \citenamefont {Fonda},\ and\ \citenamefont
  {Ghirardi}}]{Degasperis-DL-1974}%
  \BibitemOpen
  \bibfield  {author} {\bibinfo {author} {\bibfnamefont {A.}~\bibnamefont
  {Degasperis}}, \bibinfo {author} {\bibfnamefont {L.}~\bibnamefont {Fonda}},\
  and\ \bibinfo {author} {\bibfnamefont {G.~C.}\ \bibnamefont {Ghirardi}},\
  }\bibfield  {title} {\bibinfo {title} {Does the lifetime of an unstable
  system depend on the measuring apparatus?},\ }\href
  {https://doi.org/10.1007/BF02731351} {\bibfield  {journal} {\bibinfo
  {journal} {Il Nuovo Cimento A (1965-1970)}\ }\textbf {\bibinfo {volume}
  {21}},\ \bibinfo {pages} {471} (\bibinfo {year} {1974})}\BibitemShut
  {NoStop}%
\bibitem [{\citenamefont {Dhar}\ \emph
  {et~al.}(2015{\natexlab{a}})\citenamefont {Dhar}, \citenamefont {Dasgupta},\
  and\ \citenamefont {Dhar}}]{Dhar-QT-2015}%
  \BibitemOpen
  \bibfield  {author} {\bibinfo {author} {\bibfnamefont {S.}~\bibnamefont
  {Dhar}}, \bibinfo {author} {\bibfnamefont {S.}~\bibnamefont {Dasgupta}},\
  and\ \bibinfo {author} {\bibfnamefont {A.}~\bibnamefont {Dhar}},\ }\bibfield
  {title} {\bibinfo {title} {Quantum time of arrival distribution in a simple
  lattice model},\ }\href {https://doi.org/10.1088/1751-8113/48/11/115304}
  {\bibfield  {journal} {\bibinfo  {journal} {Journal of Physics A:
  Mathematical and Theoretical}\ }\textbf {\bibinfo {volume} {48}},\ \bibinfo
  {pages} {115304} (\bibinfo {year} {2015}{\natexlab{a}})}\BibitemShut
  {NoStop}%
\bibitem [{\citenamefont {Dhar}\ \emph
  {et~al.}(2015{\natexlab{b}})\citenamefont {Dhar}, \citenamefont {Dasgupta},
  \citenamefont {Dhar},\ and\ \citenamefont {Sen}}]{Dhar-DQ-2015}%
  \BibitemOpen
  \bibfield  {author} {\bibinfo {author} {\bibfnamefont {S.}~\bibnamefont
  {Dhar}}, \bibinfo {author} {\bibfnamefont {S.}~\bibnamefont {Dasgupta}},
  \bibinfo {author} {\bibfnamefont {A.}~\bibnamefont {Dhar}},\ and\ \bibinfo
  {author} {\bibfnamefont {D.}~\bibnamefont {Sen}},\ }\bibfield  {title}
  {\bibinfo {title} {Detection of a quantum particle on a lattice under
  repeated projective measurements},\ }\href
  {https://doi.org/10.1103/PhysRevA.91.062115} {\bibfield  {journal} {\bibinfo
  {journal} {Physical Review A}\ }\textbf {\bibinfo {volume} {91}},\ \bibinfo
  {pages} {062115} (\bibinfo {year} {2015}{\natexlab{b}})}\BibitemShut
  {NoStop}%
\bibitem [{\citenamefont {Friedman}\ \emph {et~al.}(2016)\citenamefont
  {Friedman}, \citenamefont {Kessler},\ and\ \citenamefont
  {Barkai}}]{Friedman-QR-2016}%
  \BibitemOpen
  \bibfield  {author} {\bibinfo {author} {\bibfnamefont {H.}~\bibnamefont
  {Friedman}}, \bibinfo {author} {\bibfnamefont {D.~A.}\ \bibnamefont
  {Kessler}},\ and\ \bibinfo {author} {\bibfnamefont {E.}~\bibnamefont
  {Barkai}},\ }\bibfield  {title} {\bibinfo {title} {Quantum renewal equation
  for the first detection time of a quantum walk},\ }\href
  {https://doi.org/10.1088/1751-8121/aa5191} {\bibfield  {journal} {\bibinfo
  {journal} {Journal of Physics A: Mathematical and Theoretical}\ }\textbf
  {\bibinfo {volume} {50}},\ \bibinfo {pages} {04LT01} (\bibinfo {year}
  {2016})}\BibitemShut {NoStop}%
\bibitem [{\citenamefont {Friedman}\ \emph {et~al.}(2017)\citenamefont
  {Friedman}, \citenamefont {Kessler},\ and\ \citenamefont
  {Barkai}}]{Friedman-QW-2017}%
  \BibitemOpen
  \bibfield  {author} {\bibinfo {author} {\bibfnamefont {H.}~\bibnamefont
  {Friedman}}, \bibinfo {author} {\bibfnamefont {D.~A.}\ \bibnamefont
  {Kessler}},\ and\ \bibinfo {author} {\bibfnamefont {E.}~\bibnamefont
  {Barkai}},\ }\bibfield  {title} {\bibinfo {title} {Quantum walks: {T}he first
  detected passage time problem},\ }\href
  {https://doi.org/10.1103/PhysRevE.95.032141} {\bibfield  {journal} {\bibinfo
  {journal} {Physical Review E}\ }\textbf {\bibinfo {volume} {95}},\ \bibinfo
  {pages} {032141} (\bibinfo {year} {2017})}\BibitemShut {NoStop}%
\bibitem [{\citenamefont {Thiel}\ \emph {et~al.}(2018)\citenamefont {Thiel},
  \citenamefont {Barkai},\ and\ \citenamefont {Kessler}}]{Thiel-FD-2018}%
  \BibitemOpen
  \bibfield  {author} {\bibinfo {author} {\bibfnamefont {F.}~\bibnamefont
  {Thiel}}, \bibinfo {author} {\bibfnamefont {E.}~\bibnamefont {Barkai}},\ and\
  \bibinfo {author} {\bibfnamefont {D.~A.}\ \bibnamefont {Kessler}},\
  }\bibfield  {title} {\bibinfo {title} {First detected arrival of a quantum
  walker on an infinite line},\ }\href
  {https://doi.org/10.1103/PhysRevLett.120.040502} {\bibfield  {journal}
  {\bibinfo  {journal} {Physical Review Letters}\ }\textbf {\bibinfo {volume}
  {120}},\ \bibinfo {pages} {040502} (\bibinfo {year} {2018})}\BibitemShut
  {NoStop}%
\bibitem [{\citenamefont {Yin}\ \emph {et~al.}(2019)\citenamefont {Yin},
  \citenamefont {Ziegler}, \citenamefont {Thiel},\ and\ \citenamefont
  {Barkai}}]{Yin-LF-2019}%
  \BibitemOpen
  \bibfield  {author} {\bibinfo {author} {\bibfnamefont {R.}~\bibnamefont
  {Yin}}, \bibinfo {author} {\bibfnamefont {K.}~\bibnamefont {Ziegler}},
  \bibinfo {author} {\bibfnamefont {F.}~\bibnamefont {Thiel}},\ and\ \bibinfo
  {author} {\bibfnamefont {E.}~\bibnamefont {Barkai}},\ }\bibfield  {title}
  {\bibinfo {title} {Large fluctuations of the first detected quantum return
  time},\ }\href {https://doi.org/10.1103/PhysRevResearch.1.033086} {\bibfield
  {journal} {\bibinfo  {journal} {Physical Review Research}\ }\textbf {\bibinfo
  {volume} {1}},\ \bibinfo {pages} {033086} (\bibinfo {year}
  {2019})}\BibitemShut {NoStop}%
\bibitem [{\citenamefont {Liu}\ \emph {et~al.}(2020)\citenamefont {Liu},
  \citenamefont {Yin}, \citenamefont {Ziegler},\ and\ \citenamefont
  {Barkai}}]{Liu-QW-2020}%
  \BibitemOpen
  \bibfield  {author} {\bibinfo {author} {\bibfnamefont {Q.}~\bibnamefont
  {Liu}}, \bibinfo {author} {\bibfnamefont {R.}~\bibnamefont {Yin}}, \bibinfo
  {author} {\bibfnamefont {K.}~\bibnamefont {Ziegler}},\ and\ \bibinfo {author}
  {\bibfnamefont {E.}~\bibnamefont {Barkai}},\ }\bibfield  {title} {\bibinfo
  {title} {Quantum walks: The mean first detected transition time},\ }\href
  {https://doi.org/10.1103/PhysRevResearch.2.033113} {\bibfield  {journal}
  {\bibinfo  {journal} {Physical Review Research}\ }\textbf {\bibinfo {volume}
  {2}},\ \bibinfo {pages} {033113} (\bibinfo {year} {2020})}\BibitemShut
  {NoStop}%
\bibitem [{\citenamefont {Thiel}\ \emph
  {et~al.}(2020{\natexlab{a}})\citenamefont {Thiel}, \citenamefont {Mualem},
  \citenamefont {Kessler},\ and\ \citenamefont {Barkai}}]{Thiel-US-2020}%
  \BibitemOpen
  \bibfield  {author} {\bibinfo {author} {\bibfnamefont {F.}~\bibnamefont
  {Thiel}}, \bibinfo {author} {\bibfnamefont {I.}~\bibnamefont {Mualem}},
  \bibinfo {author} {\bibfnamefont {D.~A.}\ \bibnamefont {Kessler}},\ and\
  \bibinfo {author} {\bibfnamefont {E.}~\bibnamefont {Barkai}},\ }\bibfield
  {title} {\bibinfo {title} {Uncertainty and symmetry bounds for the quantum
  total detection probability},\ }\href
  {https://doi.org/10.1103/PhysRevResearch.2.023392} {\bibfield  {journal}
  {\bibinfo  {journal} {Physical Review Research}\ }\textbf {\bibinfo {volume}
  {2}},\ \bibinfo {pages} {023392} (\bibinfo {year}
  {2020}{\natexlab{a}})}\BibitemShut {NoStop}%
\bibitem [{\citenamefont {Thiel}\ \emph
  {et~al.}(2020{\natexlab{b}})\citenamefont {Thiel}, \citenamefont {Mualem},
  \citenamefont {Meidan}, \citenamefont {Barkai},\ and\ \citenamefont
  {Kessler}}]{Thiel-DS-2020}%
  \BibitemOpen
  \bibfield  {author} {\bibinfo {author} {\bibfnamefont {F.}~\bibnamefont
  {Thiel}}, \bibinfo {author} {\bibfnamefont {I.}~\bibnamefont {Mualem}},
  \bibinfo {author} {\bibfnamefont {D.}~\bibnamefont {Meidan}}, \bibinfo
  {author} {\bibfnamefont {E.}~\bibnamefont {Barkai}},\ and\ \bibinfo {author}
  {\bibfnamefont {D.~A.}\ \bibnamefont {Kessler}},\ }\bibfield  {title}
  {\bibinfo {title} {Dark states of quantum search cause imperfect detection},\
  }\href {https://doi.org/10.1103/PhysRevResearch.2.043107} {\bibfield
  {journal} {\bibinfo  {journal} {Physical Review Research}\ }\textbf {\bibinfo
  {volume} {2}},\ \bibinfo {pages} {043107} (\bibinfo {year}
  {2020}{\natexlab{b}})}\BibitemShut {NoStop}%
\bibitem [{\citenamefont {Kessler}\ \emph {et~al.}(2021)\citenamefont
  {Kessler}, \citenamefont {Barkai},\ and\ \citenamefont
  {Ziegler}}]{Kessler-FD-2021}%
  \BibitemOpen
  \bibfield  {author} {\bibinfo {author} {\bibfnamefont {D.~A.}\ \bibnamefont
  {Kessler}}, \bibinfo {author} {\bibfnamefont {E.}~\bibnamefont {Barkai}},\
  and\ \bibinfo {author} {\bibfnamefont {K.}~\bibnamefont {Ziegler}},\
  }\bibfield  {title} {\bibinfo {title} {First-detection time of a quantum
  state under random probing},\ }\href
  {https://doi.org/10.1103/PhysRevA.103.022222} {\bibfield  {journal} {\bibinfo
   {journal} {Physical Review A}\ }\textbf {\bibinfo {volume} {103}},\ \bibinfo
  {pages} {022222} (\bibinfo {year} {2021})}\BibitemShut {NoStop}%
\bibitem [{\citenamefont {O'Brien}(2007)}]{OBrien-OQ-2007}%
  \BibitemOpen
  \bibfield  {author} {\bibinfo {author} {\bibfnamefont {J.~L.}\ \bibnamefont
  {O'Brien}},\ }\bibfield  {title} {\bibinfo {title} {Optical quantum
  computing},\ }\href {https://doi.org/10.1126/science.1142892} {\bibfield
  {journal} {\bibinfo  {journal} {Science}\ }\textbf {\bibinfo {volume}
  {318}},\ \bibinfo {pages} {1567} (\bibinfo {year} {2007})}\BibitemShut
  {NoStop}%
\bibitem [{\citenamefont {Childs}\ \emph {et~al.}(2013)\citenamefont {Childs},
  \citenamefont {Gosset},\ and\ \citenamefont {Webb}}]{Childs-UC-2013}%
  \BibitemOpen
  \bibfield  {author} {\bibinfo {author} {\bibfnamefont {A.~M.}\ \bibnamefont
  {Childs}}, \bibinfo {author} {\bibfnamefont {D.}~\bibnamefont {Gosset}},\
  and\ \bibinfo {author} {\bibfnamefont {Z.}~\bibnamefont {Webb}},\ }\bibfield
  {title} {\bibinfo {title} {Universal computation by multiparticle quantum
  walk},\ }\href {https://doi.org/10.1126/science.1229957} {\bibfield
  {journal} {\bibinfo  {journal} {Science}\ }\textbf {\bibinfo {volume}
  {339}},\ \bibinfo {pages} {791} (\bibinfo {year} {2013})}\BibitemShut
  {NoStop}%
\bibitem [{\citenamefont {Tichy}\ \emph {et~al.}(2010)\citenamefont {Tichy},
  \citenamefont {Tiersch}, \citenamefont {de~Melo}, \citenamefont {Mintert},\
  and\ \citenamefont {Buchleitner}}]{Tichy-ZT-2010}%
  \BibitemOpen
  \bibfield  {author} {\bibinfo {author} {\bibfnamefont {M.~C.}\ \bibnamefont
  {Tichy}}, \bibinfo {author} {\bibfnamefont {M.}~\bibnamefont {Tiersch}},
  \bibinfo {author} {\bibfnamefont {F.}~\bibnamefont {de~Melo}}, \bibinfo
  {author} {\bibfnamefont {F.}~\bibnamefont {Mintert}},\ and\ \bibinfo {author}
  {\bibfnamefont {A.}~\bibnamefont {Buchleitner}},\ }\bibfield  {title}
  {\bibinfo {title} {Zero-transmission law for multiport beam splitters},\
  }\href {https://doi.org/10.1103/PhysRevLett.104.220405} {\bibfield  {journal}
  {\bibinfo  {journal} {Physical Review Letters}\ }\textbf {\bibinfo {volume}
  {104}},\ \bibinfo {pages} {220405} (\bibinfo {year} {2010})}\BibitemShut
  {NoStop}%
\bibitem [{\citenamefont {Tichy}(2015)}]{Tichy-SP-2015}%
  \BibitemOpen
  \bibfield  {author} {\bibinfo {author} {\bibfnamefont {M.~C.}\ \bibnamefont
  {Tichy}},\ }\bibfield  {title} {\bibinfo {title} {Sampling of partially
  distinguishable bosons and the relation to the multidimensional permanent},\
  }\href {https://doi.org/10.1103/PhysRevA.91.022316} {\bibfield  {journal}
  {\bibinfo  {journal} {Physical Review A}\ }\textbf {\bibinfo {volume} {91}},\
  \bibinfo {pages} {022316} (\bibinfo {year} {2015})}\BibitemShut {NoStop}%
\bibitem [{\citenamefont {Shchesnovich}(2015)}]{Shchesnovich-PI-2015}%
  \BibitemOpen
  \bibfield  {author} {\bibinfo {author} {\bibfnamefont {V.~S.}\ \bibnamefont
  {Shchesnovich}},\ }\bibfield  {title} {\bibinfo {title} {Partial
  indistinguishability theory for multiphoton experiments in multiport
  devices},\ }\href {https://doi.org/10.1103/PhysRevA.91.013844} {\bibfield
  {journal} {\bibinfo  {journal} {Physical Review A}\ }\textbf {\bibinfo
  {volume} {91}},\ \bibinfo {pages} {013844} (\bibinfo {year}
  {2015})}\BibitemShut {NoStop}%
\bibitem [{\citenamefont {Dittel}(2019)}]{Dittel-AI-2019}%
  \BibitemOpen
  \bibfield  {author} {\bibinfo {author} {\bibfnamefont {C.}~\bibnamefont
  {Dittel}},\ }\emph {\bibinfo {title} {About the interference of many
  particles}},\ \href {https://resolver.obvsg.at/urn:nbn:at:at-ubi:1-47210}
  {Ph.D. thesis},\ \bibinfo  {school} {University of Innsbruck,
  urn:nbn:at:at-ubi:1-47210} (\bibinfo {year} {2019})\BibitemShut {NoStop}%
\bibitem [{\citenamefont {Dittel}\ \emph {et~al.}(2021)\citenamefont {Dittel},
  \citenamefont {Dufour}, \citenamefont {Weihs},\ and\ \citenamefont
  {Buchleitner}}]{Dittel-WP-2021}%
  \BibitemOpen
  \bibfield  {author} {\bibinfo {author} {\bibfnamefont {C.}~\bibnamefont
  {Dittel}}, \bibinfo {author} {\bibfnamefont {G.}~\bibnamefont {Dufour}},
  \bibinfo {author} {\bibfnamefont {G.}~\bibnamefont {Weihs}},\ and\ \bibinfo
  {author} {\bibfnamefont {A.}~\bibnamefont {Buchleitner}},\ }\bibfield
  {title} {\bibinfo {title} {Wave-particle duality of many-body quantum
  states},\ }\href {https://doi.org/10.1103/PhysRevX.11.031041} {\bibfield
  {journal} {\bibinfo  {journal} {Physical Review X}\ }\textbf {\bibinfo
  {volume} {11}},\ \bibinfo {pages} {031041} (\bibinfo {year}
  {2021})}\BibitemShut {NoStop}%
\bibitem [{\citenamefont {Minke}\ \emph {et~al.}(2021)\citenamefont {Minke},
  \citenamefont {Buchleitner},\ and\ \citenamefont {Dittel}}]{Minke-CF-2021}%
  \BibitemOpen
  \bibfield  {author} {\bibinfo {author} {\bibfnamefont {A.~M.}\ \bibnamefont
  {Minke}}, \bibinfo {author} {\bibfnamefont {A.}~\bibnamefont {Buchleitner}},\
  and\ \bibinfo {author} {\bibfnamefont {C.}~\bibnamefont {Dittel}},\
  }\bibfield  {title} {\bibinfo {title} {Characterizing four-body
  indistinguishability via symmetries},\ }\href
  {https://doi.org/10.1088/1367-2630/ac0fb1} {\bibfield  {journal} {\bibinfo
  {journal} {New Journal of Physics}\ }\textbf {\bibinfo {volume} {23}},\
  \bibinfo {pages} {073028} (\bibinfo {year} {2021})}\BibitemShut {NoStop}%
\bibitem [{\citenamefont {Lahini}\ \emph {et~al.}(2012)\citenamefont {Lahini},
  \citenamefont {Verbin}, \citenamefont {Huber}, \citenamefont {Bromberg},
  \citenamefont {Pugatch},\ and\ \citenamefont {Silberberg}}]{Lahini-QW-2012}%
  \BibitemOpen
  \bibfield  {author} {\bibinfo {author} {\bibfnamefont {Y.}~\bibnamefont
  {Lahini}}, \bibinfo {author} {\bibfnamefont {M.}~\bibnamefont {Verbin}},
  \bibinfo {author} {\bibfnamefont {S.~D.}\ \bibnamefont {Huber}}, \bibinfo
  {author} {\bibfnamefont {Y.}~\bibnamefont {Bromberg}}, \bibinfo {author}
  {\bibfnamefont {R.}~\bibnamefont {Pugatch}},\ and\ \bibinfo {author}
  {\bibfnamefont {Y.}~\bibnamefont {Silberberg}},\ }\bibfield  {title}
  {\bibinfo {title} {Quantum walk of two interacting bosons},\ }\href
  {https://doi.org/10.1103/PhysRevA.86.011603} {\bibfield  {journal} {\bibinfo
  {journal} {Physical Review A}\ }\textbf {\bibinfo {volume} {86}},\ \bibinfo
  {pages} {011603} (\bibinfo {year} {2012})}\BibitemShut {NoStop}%
\bibitem [{\citenamefont {Ahlbrecht}\ \emph {et~al.}(2012)\citenamefont
  {Ahlbrecht}, \citenamefont {Alberti}, \citenamefont {Meschede}, \citenamefont
  {Scholz}, \citenamefont {Werner},\ and\ \citenamefont
  {Werner}}]{Ahlbrecht-MB-2012}%
  \BibitemOpen
  \bibfield  {author} {\bibinfo {author} {\bibfnamefont {A.}~\bibnamefont
  {Ahlbrecht}}, \bibinfo {author} {\bibfnamefont {A.}~\bibnamefont {Alberti}},
  \bibinfo {author} {\bibfnamefont {D.}~\bibnamefont {Meschede}}, \bibinfo
  {author} {\bibfnamefont {V.~B.}\ \bibnamefont {Scholz}}, \bibinfo {author}
  {\bibfnamefont {A.~H.}\ \bibnamefont {Werner}},\ and\ \bibinfo {author}
  {\bibfnamefont {R.~F.}\ \bibnamefont {Werner}},\ }\bibfield  {title}
  {\bibinfo {title} {Molecular binding in interacting quantum walks},\ }\href
  {https://doi.org/10.1088/1367-2630/14/7/073050} {\bibfield  {journal}
  {\bibinfo  {journal} {New Journal of Physics}\ }\textbf {\bibinfo {volume}
  {14}},\ \bibinfo {pages} {073050} (\bibinfo {year} {2012})}\BibitemShut
  {NoStop}%
\bibitem [{\citenamefont {Preiss}\ \emph {et~al.}(2015)\citenamefont {Preiss},
  \citenamefont {Ma}, \citenamefont {Tai}, \citenamefont {Lukin}, \citenamefont
  {Rispoli}, \citenamefont {Zupancic}, \citenamefont {Lahini}, \citenamefont
  {Islam},\ and\ \citenamefont {Greiner}}]{Preiss-SC-2015}%
  \BibitemOpen
  \bibfield  {author} {\bibinfo {author} {\bibfnamefont {P.~M.}\ \bibnamefont
  {Preiss}}, \bibinfo {author} {\bibfnamefont {R.}~\bibnamefont {Ma}}, \bibinfo
  {author} {\bibfnamefont {M.~E.}\ \bibnamefont {Tai}}, \bibinfo {author}
  {\bibfnamefont {A.}~\bibnamefont {Lukin}}, \bibinfo {author} {\bibfnamefont
  {M.}~\bibnamefont {Rispoli}}, \bibinfo {author} {\bibfnamefont
  {P.}~\bibnamefont {Zupancic}}, \bibinfo {author} {\bibfnamefont
  {Y.}~\bibnamefont {Lahini}}, \bibinfo {author} {\bibfnamefont
  {R.}~\bibnamefont {Islam}},\ and\ \bibinfo {author} {\bibfnamefont
  {M.}~\bibnamefont {Greiner}},\ }\bibfield  {title} {\bibinfo {title}
  {Strongly correlated quantum walks in optical lattices},\ }\href
  {https://doi.org/10.1126/science.1260364} {\bibfield  {journal} {\bibinfo
  {journal} {Science}\ }\textbf {\bibinfo {volume} {347}},\ \bibinfo {pages}
  {1229} (\bibinfo {year} {2015})}\BibitemShut {NoStop}%
\bibitem [{\citenamefont {Cai}\ \emph {et~al.}(2021)\citenamefont {Cai},
  \citenamefont {Yang}, \citenamefont {Shi}, \citenamefont {Lee}, \citenamefont
  {Andrei},\ and\ \citenamefont {Guan}}]{Cai-MQ-2021}%
  \BibitemOpen
  \bibfield  {author} {\bibinfo {author} {\bibfnamefont {X.}~\bibnamefont
  {Cai}}, \bibinfo {author} {\bibfnamefont {H.}~\bibnamefont {Yang}}, \bibinfo
  {author} {\bibfnamefont {H.-L.}\ \bibnamefont {Shi}}, \bibinfo {author}
  {\bibfnamefont {C.}~\bibnamefont {Lee}}, \bibinfo {author} {\bibfnamefont
  {N.}~\bibnamefont {Andrei}},\ and\ \bibinfo {author} {\bibfnamefont {X.-W.}\
  \bibnamefont {Guan}},\ }\bibfield  {title} {\bibinfo {title} {Multiparticle
  quantum walks and fisher information in one-dimensional lattices},\ }\href
  {https://doi.org/10.1103/PhysRevLett.127.100406} {\bibfield  {journal}
  {\bibinfo  {journal} {Physical Review Letters}\ }\textbf {\bibinfo {volume}
  {127}},\ \bibinfo {pages} {100406} (\bibinfo {year} {2021})}\BibitemShut
  {NoStop}%
\bibitem [{\citenamefont {Poulios}\ \emph {et~al.}(2014)\citenamefont
  {Poulios}, \citenamefont {Keil}, \citenamefont {Fry}, \citenamefont
  {Meinecke}, \citenamefont {Matthews}, \citenamefont {Politi}, \citenamefont
  {Lobino}, \citenamefont {Gr\"afe}, \citenamefont {Heinrich}, \citenamefont
  {Nolte}, \citenamefont {Szameit},\ and\ \citenamefont
  {O'Brien}}]{Poulios-QW-2014}%
  \BibitemOpen
  \bibfield  {author} {\bibinfo {author} {\bibfnamefont {K.}~\bibnamefont
  {Poulios}}, \bibinfo {author} {\bibfnamefont {R.}~\bibnamefont {Keil}},
  \bibinfo {author} {\bibfnamefont {D.}~\bibnamefont {Fry}}, \bibinfo {author}
  {\bibfnamefont {J.~D.~A.}\ \bibnamefont {Meinecke}}, \bibinfo {author}
  {\bibfnamefont {J.~C.~F.}\ \bibnamefont {Matthews}}, \bibinfo {author}
  {\bibfnamefont {A.}~\bibnamefont {Politi}}, \bibinfo {author} {\bibfnamefont
  {M.}~\bibnamefont {Lobino}}, \bibinfo {author} {\bibfnamefont
  {M.}~\bibnamefont {Gr\"afe}}, \bibinfo {author} {\bibfnamefont
  {M.}~\bibnamefont {Heinrich}}, \bibinfo {author} {\bibfnamefont
  {S.}~\bibnamefont {Nolte}}, \bibinfo {author} {\bibfnamefont
  {A.}~\bibnamefont {Szameit}},\ and\ \bibinfo {author} {\bibfnamefont {J.~L.}\
  \bibnamefont {O'Brien}},\ }\bibfield  {title} {\bibinfo {title} {Quantum
  walks of correlated photon pairs in two-dimensional waveguide arrays},\
  }\href {https://doi.org/10.1103/PhysRevLett.112.143604} {\bibfield  {journal}
  {\bibinfo  {journal} {Physical Review Letters}\ }\textbf {\bibinfo {volume}
  {112}},\ \bibinfo {pages} {143604} (\bibinfo {year} {2014})}\BibitemShut
  {NoStop}%
\bibitem [{\citenamefont {Crespi}\ \emph {et~al.}(2015)\citenamefont {Crespi},
  \citenamefont {Sansoni}, \citenamefont {Valle}, \citenamefont {Ciamei},
  \citenamefont {Ramponi}, \citenamefont {Sciarrino}, \citenamefont {Mataloni},
  \citenamefont {Longhi},\ and\ \citenamefont {Osellame}}]{Crespi-PS-2015}%
  \BibitemOpen
  \bibfield  {author} {\bibinfo {author} {\bibfnamefont {A.}~\bibnamefont
  {Crespi}}, \bibinfo {author} {\bibfnamefont {L.}~\bibnamefont {Sansoni}},
  \bibinfo {author} {\bibfnamefont {G.~D.}\ \bibnamefont {Valle}}, \bibinfo
  {author} {\bibfnamefont {A.}~\bibnamefont {Ciamei}}, \bibinfo {author}
  {\bibfnamefont {R.}~\bibnamefont {Ramponi}}, \bibinfo {author} {\bibfnamefont
  {F.}~\bibnamefont {Sciarrino}}, \bibinfo {author} {\bibfnamefont
  {P.}~\bibnamefont {Mataloni}}, \bibinfo {author} {\bibfnamefont
  {S.}~\bibnamefont {Longhi}},\ and\ \bibinfo {author} {\bibfnamefont
  {R.}~\bibnamefont {Osellame}},\ }\bibfield  {title} {\bibinfo {title}
  {Particle statistics affects quantum decay and {F}ano interference},\ }\href
  {https://doi.org/10.1103/PhysRevLett.114.090201} {\bibfield  {journal}
  {\bibinfo  {journal} {Physical Review Letters}\ }\textbf {\bibinfo {volume}
  {114}},\ \bibinfo {pages} {090201} (\bibinfo {year} {2015})}\BibitemShut
  {NoStop}%
\bibitem [{\citenamefont {Ehrhardt}\ \emph {et~al.}(2021)\citenamefont
  {Ehrhardt}, \citenamefont {Keil}, \citenamefont {Maczewsky}, \citenamefont
  {Dittel}, \citenamefont {Heinrich},\ and\ \citenamefont
  {Szameit}}]{Ehrhardt-EC-2021}%
  \BibitemOpen
  \bibfield  {author} {\bibinfo {author} {\bibfnamefont {M.}~\bibnamefont
  {Ehrhardt}}, \bibinfo {author} {\bibfnamefont {R.}~\bibnamefont {Keil}},
  \bibinfo {author} {\bibfnamefont {L.~J.}\ \bibnamefont {Maczewsky}}, \bibinfo
  {author} {\bibfnamefont {C.}~\bibnamefont {Dittel}}, \bibinfo {author}
  {\bibfnamefont {M.}~\bibnamefont {Heinrich}},\ and\ \bibinfo {author}
  {\bibfnamefont {A.}~\bibnamefont {Szameit}},\ }\bibfield  {title} {\bibinfo
  {title} {Exploring complex graphs using three-dimensional quantum walks of
  correlated photons},\ }\href {https://doi.org/10.1126/sciadv.abc5266}
  {\bibfield  {journal} {\bibinfo  {journal} {Science Advances}\ }\textbf
  {\bibinfo {volume} {7}},\ \bibinfo {pages} {eabc5266} (\bibinfo {year}
  {2021})}\BibitemShut {NoStop}%
\bibitem [{\citenamefont {Brunner}\ \emph {et~al.}(2022)\citenamefont
  {Brunner}, \citenamefont {Buchleitner},\ and\ \citenamefont
  {Dufour}}]{Brunner-MB-2022}%
  \BibitemOpen
  \bibfield  {author} {\bibinfo {author} {\bibfnamefont {E.}~\bibnamefont
  {Brunner}}, \bibinfo {author} {\bibfnamefont {A.}~\bibnamefont
  {Buchleitner}},\ and\ \bibinfo {author} {\bibfnamefont {G.}~\bibnamefont
  {Dufour}},\ }\bibfield  {title} {\bibinfo {title} {Many-body coherence and
  entanglement probed by randomized correlation measurements},\ }\href
  {https://doi.org/10.1103/PhysRevResearch.4.043101} {\bibfield  {journal}
  {\bibinfo  {journal} {Physical Review Research}\ }\textbf {\bibinfo {volume}
  {4}},\ \bibinfo {pages} {043101} (\bibinfo {year} {2022})}\BibitemShut
  {NoStop}%
\bibitem [{Note1()}]{Note1}%
  \BibitemOpen
  \bibinfo {note} {Note that diagonalizable complex valued matrices form a
  dense subset of the set of $n\times n$ complex matrices. Hence, almost all
  $n\times n$ complex matrices are diagonalizable, and those which are not can
  be approximated to arbitrary precision by a diagonalizable matrix \cite
  {Horn-MA-2013}.}\BibitemShut {Stop}%
\bibitem [{\citenamefont {Hildebrandt}(2006)}]{Hildebrandt-Ana-2006}%
  \BibitemOpen
  \bibfield  {author} {\bibinfo {author} {\bibfnamefont {S.}~\bibnamefont
  {Hildebrandt}},\ }\href
  {https://doi.org/https://doi.org/10.1007/3-540-29285-3} {\emph {\bibinfo
  {title} {Analysis 1}}},\ \bibinfo {edition} {2nd}\ ed.\ (\bibinfo
  {publisher} {Springer, Berlin, Heidelberg},\ \bibinfo {year}
  {2006})\BibitemShut {NoStop}%
\bibitem [{Note2()}]{Note2}%
  \BibitemOpen
  \bibinfo {note} {Note that the more the particles initially bunch, the less
  can there be many-particle interference \cite
  {Dittel-AI-2019,Dittel-WP-2021}. Hence, we choose one particle per site to
  allow for strong interference effects.}\BibitemShut {Stop}%
\bibitem [{Note3()}]{Note3}%
  \BibitemOpen
  \bibinfo {note} {A multiset is a generalization of a set, where multiple
  instances of the same element are allowed.}\BibitemShut {Stop}%
\bibitem [{Note4()}]{Note4}%
  \BibitemOpen
  \bibinfo {note} {A \protect \textit {transversal} of a collection of sets
  $B_1,\protect \dots ,B_R$ is a set of $R$ elements, which contains exactly
  one element of each set $B_1,\protect \dots ,B_R$. For $H$ a subgroup of the
  group $G$, the \protect \textit {right transversal} of $H$ in $G$ is a
  transversal of the set of distinct right cosets of $H$ in $G$. The right
  coset of $H$ in $G$ with respect to $\pi \in H$ is $H\pi =\protect \{ \xi \pi
  \ | \ \xi \in H \protect \}$ \cite {Baumslag-SO-1968}.}\BibitemShut {Stop}%
\bibitem [{\citenamefont {Carolan}\ \emph {et~al.}(2014)\citenamefont
  {Carolan}, \citenamefont {Meinecke}, \citenamefont {Shadbolt}, \citenamefont
  {Russell}, \citenamefont {Ismail}, \citenamefont {W{\"o}rhoff}, \citenamefont
  {Rudolph}, \citenamefont {Thompson}, \citenamefont {O'Brien}, \citenamefont
  {Matthews},\ and\ \citenamefont {Laing}}]{Carolan-OE-2014}%
  \BibitemOpen
  \bibfield  {author} {\bibinfo {author} {\bibfnamefont {J.}~\bibnamefont
  {Carolan}}, \bibinfo {author} {\bibfnamefont {J.~D.~A.}\ \bibnamefont
  {Meinecke}}, \bibinfo {author} {\bibfnamefont {P.~J.}\ \bibnamefont
  {Shadbolt}}, \bibinfo {author} {\bibfnamefont {N.~J.}\ \bibnamefont
  {Russell}}, \bibinfo {author} {\bibfnamefont {N.}~\bibnamefont {Ismail}},
  \bibinfo {author} {\bibfnamefont {K.}~\bibnamefont {W{\"o}rhoff}}, \bibinfo
  {author} {\bibfnamefont {T.}~\bibnamefont {Rudolph}}, \bibinfo {author}
  {\bibfnamefont {M.~G.}\ \bibnamefont {Thompson}}, \bibinfo {author}
  {\bibfnamefont {J.~L.}\ \bibnamefont {O'Brien}}, \bibinfo {author}
  {\bibfnamefont {J.~C.~F.}\ \bibnamefont {Matthews}},\ and\ \bibinfo {author}
  {\bibfnamefont {A.}~\bibnamefont {Laing}},\ }\bibfield  {title} {\bibinfo
  {title} {On the experimental verification of quantum complexity in linear
  optics},\ }\href {https://doi.org/10.1038/nphoton.2014.152} {\bibfield
  {journal} {\bibinfo  {journal} {Nature Photonics}\ }\textbf {\bibinfo
  {volume} {8}},\ \bibinfo {pages} {621} (\bibinfo {year} {2014})}\BibitemShut
  {NoStop}%
\bibitem [{\citenamefont {Shchesnovich}(2016)}]{Shchesnovich-UG-2016}%
  \BibitemOpen
  \bibfield  {author} {\bibinfo {author} {\bibfnamefont {V.~S.}\ \bibnamefont
  {Shchesnovich}},\ }\bibfield  {title} {\bibinfo {title} {Universality of
  generalized bunching and efficient assessment of boson sampling},\ }\href
  {https://doi.org/10.1103/PhysRevLett.116.123601} {\bibfield  {journal}
  {\bibinfo  {journal} {Physical Review Letters}\ }\textbf {\bibinfo {volume}
  {116}},\ \bibinfo {pages} {123601} (\bibinfo {year} {2016})}\BibitemShut
  {NoStop}%
\bibitem [{Note5()}]{Note5}%
  \BibitemOpen
  \bibinfo {note} {Note that an increasing first detection time for increasing
  symmetry of the many-body state is no generally valid trend. Instead, it is
  an artifact of the chosen lattice Hamiltonian. For example, for a
  one-dimensional linear lattice with three sites and periodic boundary
  conditions (i.e. for a ring), and for the detection of exactly one particle
  $(=1)$, we find sampling times for which the first detection time increases
  with decreasing symmetry of the two-body state \cite
  {Neubrand-FD-2020}.}\BibitemShut {Stop}%
\bibitem [{\citenamefont {Allcock}(1969)}]{Allock-TA-1969}%
  \BibitemOpen
  \bibfield  {author} {\bibinfo {author} {\bibfnamefont {G.~R.}\ \bibnamefont
  {Allcock}},\ }\bibfield  {title} {\bibinfo {title} {The time of arrival in
  quantum mechanics ii. the individual measurement},\ }\href
  {https://doi.org/https://doi.org/10.1016/0003-4916(69)90252-8} {\bibfield
  {journal} {\bibinfo  {journal} {Annals of Physics}\ }\textbf {\bibinfo
  {volume} {53}},\ \bibinfo {pages} {286} (\bibinfo {year} {1969})}\BibitemShut
  {NoStop}%
\bibitem [{\citenamefont {Mayer}\ \emph {et~al.}(2011)\citenamefont {Mayer},
  \citenamefont {Tichy}, \citenamefont {Mintert}, \citenamefont {Konrad},\ and\
  \citenamefont {Buchleitner}}]{Mayer-CS-2011}%
  \BibitemOpen
  \bibfield  {author} {\bibinfo {author} {\bibfnamefont {K.}~\bibnamefont
  {Mayer}}, \bibinfo {author} {\bibfnamefont {M.~C.}\ \bibnamefont {Tichy}},
  \bibinfo {author} {\bibfnamefont {F.}~\bibnamefont {Mintert}}, \bibinfo
  {author} {\bibfnamefont {T.}~\bibnamefont {Konrad}},\ and\ \bibinfo {author}
  {\bibfnamefont {A.}~\bibnamefont {Buchleitner}},\ }\bibfield  {title}
  {\bibinfo {title} {Counting statistics of many-particle quantum walks},\
  }\href {https://doi.org/10.1103/PhysRevA.83.062307} {\bibfield  {journal}
  {\bibinfo  {journal} {Physical Review A}\ }\textbf {\bibinfo {volume} {83}},\
  \bibinfo {pages} {062307} (\bibinfo {year} {2011})}\BibitemShut {NoStop}%
\bibitem [{\citenamefont {Horn}\ and\ \citenamefont
  {Johnson}(2013)}]{Horn-MA-2013}%
  \BibitemOpen
  \bibfield  {author} {\bibinfo {author} {\bibfnamefont {R.~A.}\ \bibnamefont
  {Horn}}\ and\ \bibinfo {author} {\bibfnamefont {C.~R.}\ \bibnamefont
  {Johnson}},\ }\href@noop {} {\emph {\bibinfo {title} {Matrix Analysis}}},\
  \bibinfo {edition} {2nd}\ ed.\ (\bibinfo  {publisher} {Cambridge University
  Press},\ \bibinfo {address} {Cambridge; New York},\ \bibinfo {year}
  {2013})\BibitemShut {NoStop}%
\bibitem [{\citenamefont {Baumslag}\ and\ \citenamefont
  {Chandler}(1968)}]{Baumslag-SO-1968}%
  \BibitemOpen
  \bibfield  {author} {\bibinfo {author} {\bibfnamefont {B.}~\bibnamefont
  {Baumslag}}\ and\ \bibinfo {author} {\bibfnamefont {B.}~\bibnamefont
  {Chandler}},\ }\href@noop {} {\emph {\bibinfo {title} {Schaum's outline of
  theory and problems of group theory}}},\ Schaum's outline series\ (\bibinfo
  {publisher} {McGraw-Hill},\ \bibinfo {address} {New York},\ \bibinfo {year}
  {1968})\ \bibinfo {note} {cover title: Theory and problems of group
  theory.}\BibitemShut {Stop}%
\bibitem [{\citenamefont {Neubrand}(2020)}]{Neubrand-FD-2020}%
  \BibitemOpen
  \bibfield  {author} {\bibinfo {author} {\bibfnamefont {N.}~\bibnamefont
  {Neubrand}},\ }\bibfield  {title} {\bibinfo {title} {First detection time
  statistics of many partially distinguishable particles},\ }\bibfield
  {journal} {\bibinfo  {journal} {{B.Sc.} thesis,
  {A}lbert-{L}udwigs-{U}niversit{\"a}t {F}reiburg,
  urn:nbn:de:bsz:25-freidok-2332555}\ }\href
  {https://doi.org/10.6094/UNIFR/233255} {10.6094/UNIFR/233255} (\bibinfo
  {year} {2020})\BibitemShut {NoStop}%
\end{thebibliography}

%

\end{document}